%% file: RPA_Sternheimer.tex
\documentclass[journal=jctcce,manuscript=article]{achemso}

\usepackage{chemformula} 
\usepackage[T1]{fontenc} 

\usepackage{graphicx}

\usepackage{siunitx}
\usepackage{bm}
\usepackage{float}
\usepackage{subfig}
\usepackage[T1]{fontenc}
\usepackage{amsmath,amssymb}
\usepackage{dcolumn}
\usepackage{epsfig}
\usepackage{longtable}
\usepackage{multirow}
\usepackage{color}
\usepackage{booktabs}
\usepackage{CJK}
\usepackage{indentfirst}
\usepackage[colorlinks=true, citecolor=black, urlcolor=black,linkcolor=black]{hyperref}     

\author{Hao Peng}
\affiliation{Beijing National Laboratory for Condensed Matter Physics, Institute of Physics, Chinese Academy of Sciences, Beijing 100190, China}
\alsoaffiliation{University of Chinese Academy of Sciences, Beijing 100049, China}
\author{Sixian Yang}
\affiliation{Key Laboratory of Quantum Information, University of Science and Technology of China, Hefei 230026, China}
\alsoaffiliation{Beijing National Laboratory for Condensed Matter Physics, Institute of Physics, Chinese Academy of Sciences, Beijing 100190, China}
\author{Hong Jiang}
\affiliation{Beijing National Laboratory for Molecular Sciences, College of Chemistry and Molecular Engineering, Peking University, 100871 Beijing, China}
\email{jianghchem@pku.edu.cn}
\author{Hongming Weng}
\affiliation{Beijing National Laboratory for Condensed Matter Physics, Institute of Physics, Chinese Academy of Sciences, Beijing 100190, China}
\alsoaffiliation{Songshan Lake Materials Laboratory, Dongguan 523808, Guangdong, China}
\email{hmweng@iphy.ac.cn}
\author{Xinguo Ren}
\email{renxg@iphy.ac.cn}
\affiliation{Beijing National Laboratory for Condensed Matter Physics, Institute of Physics, Chinese Academy of Sciences, Beijing 100190, China}
\alsoaffiliation{Songshan Lake Materials Laboratory, Dongguan 523808, Guangdong, China}

\title{Basis-set-error-free RPA correlation energies for atoms based on the Sternheimer equation}

\begin{document}
\include{newcommands}

 \begin{tocentry}




  \centering
  \includegraphics[width=0.8\textwidth]{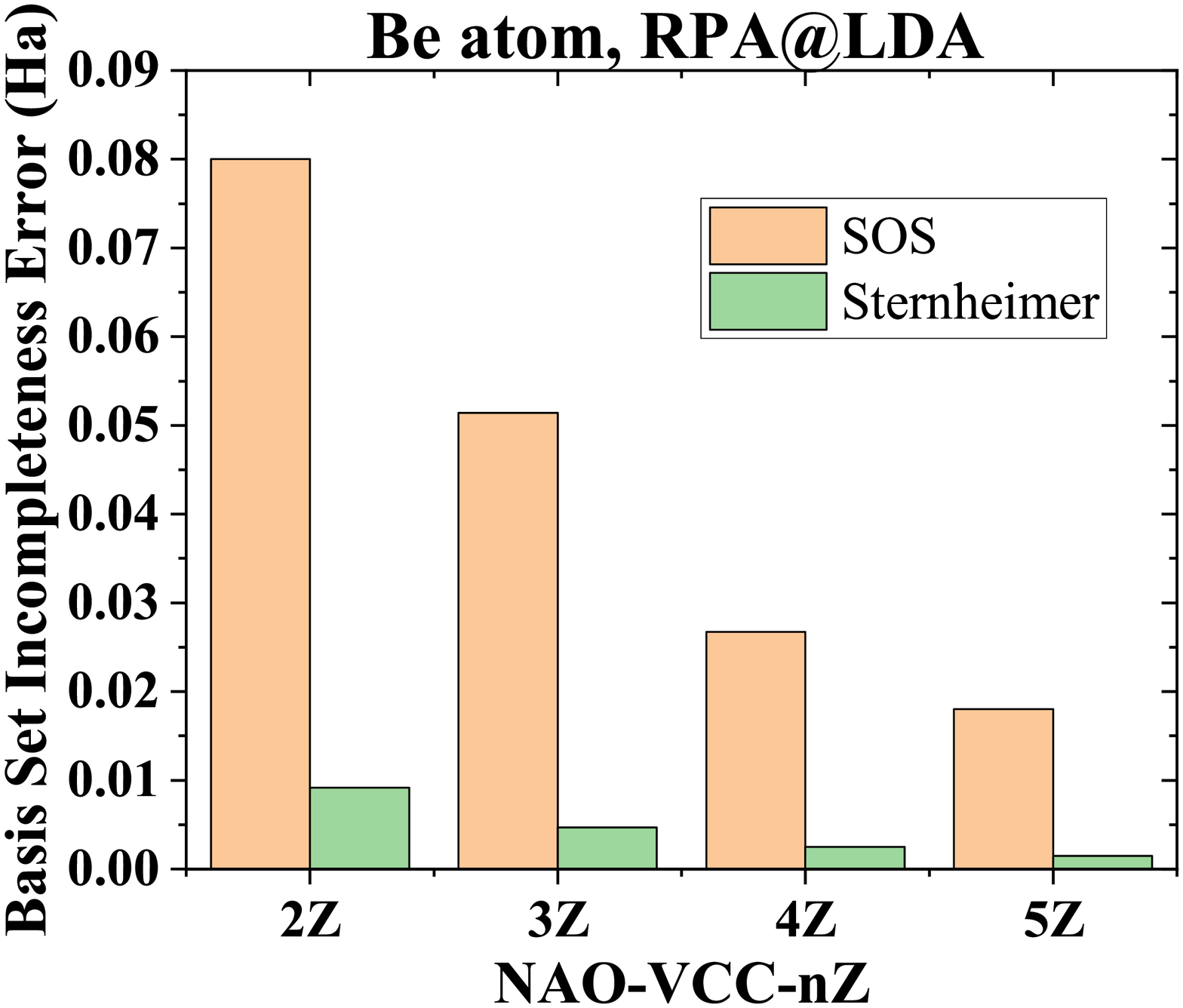}
  
\end{tocentry}

\begin{abstract}
The finite basis set errors for all-electron random-phase approximation (RPA) 
correlation energy calculations are analyzed for isolated atomic systems. We show that, 
within the resolution-of-identity (RI) RPA framework, the major source of the basis set errors is the incompleteness of the
single-particle atomic orbitals used to expand the Kohn-Sham eigenstates, instead of the auxiliary basis set (ABS) to 
represent the density response function $\chi^0$ and the bare Coulomb operator $v$. By solving the Sternheimer equation for
the first-order wave function on a dense radial grid, we are able to eliminate the major error -- the incompleteness error of
the single-particle atomic basis set -- for atomic RPA calculations. The error stemming from a finite ABS can be readily rendered vanishingly small
by increasing the size of the ABS, or by iteratively determining the eigenmodes of the $\chi^0 v$ operator. The variational property of the RI-RPA correlation energy can be further exploited to optimize
the ABS in order to achieve a fast convergence of the RI-RPA correlation energy. These numerical techniques enable us to obtain basis-set-error-free RPA correlation energies for atoms,
and in this work such energies for atoms from H to Kr are presented.
The implications of the numerical techniques developed in the present work for addressing the basis set issue for molecules and solids are discussed.
\end{abstract}

\section{Introduction}

First-principles electronic-structure calculations play a pivotal role in studying molecular and materials' properties. In order to achieve
predictive accuracy of such calculations, there have been continuing efforts devoted to developing novel electronic-structure methods and
algorithms to capture many-electron correlations at affordable cost. However, in addition to the intrinsic error of the
 electron-structure models, there is
another source of error that limits the numerical accuracy of electronic structure calculations, stemming from 
a limited resolution in representing the wavefunctions and/or other
key quantities, such as the density response functions. This latter type of error 
is commonly known as basis set incompleteness error (BSIE). In computational 
chemistry, the molecular orbitals are usually expanded in terms of atomic orbitals (AOs), 
such as the popular Gaussian-type orbitals (GTOs) or numerical 
atomic orbitals (NAOs). Compared to the plane-wave basis sets, 
a comparable numerical precision can be achieved for AO basis sets of much smaller size. However, such basis sets are generally non-orthogonal for polyatomic systems and the BSIE cannot be systematically eliminated
by simply increasing the basis set size. In practice, by employing the ``correlation consistent'' GTO \cite{Dunning:1989} or NAO basis sets \cite{IgorZhang/etal:2013},  empirical rules have been developed for extrapolating the results obtained using finite AO
basis sets to the 
so-called ``compete basis set'' (CBS) limit \cite{Helgaker/etal:1997}. However, a rigorous assessment of the reliability of such CBS limit 
is still difficult, without independent reference data to compare with.
The slow convergence of correlated methods with respect to the single-particle basis sets (SPBS) lies in the difficulty of describing the virtual space, or equivalently the unoccupied molecular orbitals. In fact, the involvement of virtual orbitals in the formalism comes from 
the expansion of the variation
of the (many- or single-particle) wavefunctions in terms of the entire set of eigenfunctions of the non-perturbed zeroth-order 
Hamiltonian. Thus, if one can directly compute the perturbed wavefunctions, one should be able to remove the explicit dependence of 
the virtual orbitals in the formalism. 

In recent years, the random phase approximation (RPA) as a ground-state total energy method has attracted much interest in 
computational chemistry and materials science \cite{Eshuis/Bates/Furche:2012,Ren/etal:2012b,Zhang/Ren:2023}. The key ingredient 
in RPA correlation energy calculations is the Kohn-Sham (KS) response function $\chi^0$, which describes how the electron density 
varies upon a disturbance in the KS effective potential.
The variation of the electron density $\delta n(\bfr)$ is determined by the change of occupied KS orbitals $\delta \psi_n(\bfr)$, 
which is usually expanded in term of the full
spectrum of the eigenfunctions of the unperturbed KS Hamiltonian. This results in the familiar sum-over-states (SOS) 
expression of the KS response function, requiring an infinite summation over unoccupied states.
Consequently, similar to other correlated methods, it becomes a real challenge to obtain fully converged RPA correlation energies
in terms of basis set expansion, regardless of the specific basis type -- plane waves or atomic orbitals. In practical RPA calculations, 
one often has to eliminate the contributions 
from core electrons via the pseudopotential scheme or the frozen-core approximation, 
and employs valance correlation consistent basis sets
which allow for extrapolations. Furthermore, since the energy differences such as the cohesive energies or adsorption energies are of
real interest, the numerical precision of practical correlated calculations relies on the cancellation of the BSIE in the correlation energies of the whole system 
 and of its fragments. Nevertheless, obtaining reference results that are free of BSIE for prototypical
systems is of great interest for both both benchmark purpose and for guiding the development of more efficient AO basis sets. Recently, such effort has
been taken to benchmark the absolute ground-state total energy of DFT calculations for atoms and small molecules \cite{Jensen/etal:2017}, 
although the situation there is less challenging as only occupied states are required.


An alternative way of obtaining the KS response function is to determine directly the density changes $\delta n(\bfr)$, or equivalently
the changes of KS orbitals $\delta \psi_n(\bfr)$ in response to trial perturbative potentials $\delta V(\bfr)$. 
The equation governing the relationship between
$\delta \psi_n(\bfr)$ and $\delta V(\bfr)$ to the linear order is known as the Sternheimer equation \cite{Sternheimer:1951,sternheimer1954electronic}. 
Obviously, the numerical precision
of the solution of such an equation depends on how $\delta \psi_n(\bfr)$ is discretized. If one chooses to represent $\delta \psi_n(\bfr)$ in
terms of the same basis set that is employed to expand the unperturbed KS orbitals $\{\psi_n(\bfr)\}$, one would obtain exactly the same results
as yielded by the SOS approach. However, if the spatial resolution of $\delta \psi_n(\bfr)$ goes beyond what can be characterized 
by the original SPBS, one can then obtain results that are better converged. Employing the Sternheimer-like equations to solve the response problem has
a long history \cite{Sternheimer:1951,sternheimer1957electronic,sternheimer1970quadrupole,mahan1980modified,zangwill1980density,mahan1982van}. It
was originally used by Sternheimer to compute nuclear quadrupole moments \cite{Sternheimer:1951}, as well as
the electronic dipole and quadrupole polarizabilities of atomic ions \cite{sternheimer1954electronic,sternheimer1957electronic} and
alkali atoms \cite{sternheimer1970quadrupole}. A self-consistent version of the Sternheimer equation was used by Mahan to compute the static and dynamic polarizabilities \cite{mahan1980modified}, and van der Waals (vdW)
coefficients \cite{mahan1982van} of closed-shell ions, where the exchange-correlation potential was treated by 
the local-density approximation (LDA) of DFT. Similar scheme has been used by Zangwill and Soven to compute the photoabsorption cross section
of rare-gas atoms \cite{zangwill1980density}.  More recently, a frequency-dependent Sternheimer equation has been used 
to compute the dynamical hyperpolarizability of molecules \cite{Andrade/etal:2007} within the formalism of 
time-dependent DFT (TDDFT). In quantum chemistry, similar equations based on the Hartree-Fock (HF) approximation are termed as
coupled perturbed HF equations \cite{Gerratt/Mills:1968}, which has been used to compute the force constants and dipole-moment derivatives of molecules. In solid-state community, the Sternheimer-like approach formulated as a linear response
of KS-DFT equations is often called density-functional perturbation theory (DFPT), which has been routinely used
to compute lattice vibrations and electron-phonon couplings, in particular in the pseudopotential plane-wave framework
\cite{Baroni/etal:1987,Baroni/etal:2001,Gonze:1995}.

More recently, the Sternheimer-type approach, or DFPT, has been employed in the RPA correlation energy \cite{wilson2008efficient,wilson2009iterative,nguyen2009efficient} and $GW$ self-energy calculations \cite{Umari/Stenuit/Baroni:2009,Umari/Stenuit/Baroni:2010,Giustino/Cohen/Louie:2010,Nguyen/etal:2012} with the intention of removing the
empty states in the formulation. Reduced prefactors of the computational cost and improved quality of convergence were reported in
these studies. These ``without empty states" implementations are mostly carried out in the pseudopotential plane-wave framework, utilizing
the existing infrastructure and experience of DFPT there \cite{Baroni/etal:2001}. It should be noted that, however, the results yielded
by formalisms without empty states is not equivalent to the CBS limit. The quality of the basis set convergence of such calculations is ultimately
determined by how the first-order wavefunctions or density are discretized, i.e., the energy cutoff of the plane waves to represent the
first-order quantities. Besides the efforts in the pseudopotential plane-wave community, another interesting development was reported in
the framework of linearized augmented plane-wave (LAPW) basis set framework \cite{betzinger2012precise,betzinger2013precise,betzinger2015precise}. 
Specifically, a radial Sternheimer equation was solved by numerical integration to arrive at an incomplete basis-set correction (IBC) to
the density response that is not captured in the Hilbert space spanned by the original LAPW basis set. Application of the IBC to the
optimized effective potential (OEP) \cite{betzinger2012precise,betzinger2013precise} and RPA correlation energy calculations 
\cite{betzinger2015precise} showed that the Sternheimer approach can significantly speed up the basis set convergence of such calculations.

In this work, the Sternheimer approach is used as a powerful numerical technique to obtain highly converged RPA correlation energy 
within an AO basis set framework. Specifically we employ the resolution of identity (RI) formulation of RPA. within which, in addition to the
standard AO basis set, an auxiliary basis set (ABS) is invoked to represent the two-particle quantities -- here the bare Coulomb interaction $v$ and 
non-interacting KS response function $\chi^0$. For numerical
simplicity, the present study is restricted to isolated atoms, whereby the Sternheimer equation is reduced to an one-dimensional radial
differential equation which can be solved accurately by numerical integration on a dense grid. 
We show that, for a given ABS, the RI-RPA correlation energy
is highly converged, and the BSIE stemming from the SPBS is vanishing small. This enables one to exploit the variational property of the
RI-RPA correlation energy with respect to the ABS, from which one can optimize the ABS by minimizing the RI-RPA correlation energy. 
Last but not least, by analyzing the difference of the density response obtained in terms of the Sternheimer approach and the SOS approach, we can
gain insights about the deficiency of the currently used basis sets, which provides guidance for generating efficient 
complementary basis functions to improve the basis set convergence quality within the standard SOS scheme.

The paper is organized as follows. In Sec.~\ref{sec:theory} the theoretical formulations is presented, where we review and compare the
formalism for calculating the RI-RPA correlation energies in terms of the SOS and Sternheimer schemes. This is followed by a brief
discussion of the actual implementation in Sec.~\ref{sec:imple}. Section~\ref{sec:results} presents the major results obtained in this work,
including a comparison of the basis set convergence behavior between the SOS and Sternheimer schemes, an error analysis of the first-order
density as yielded by the usual SOS approach, and the performance of an ABS optimization scheme based on the variational property of
the RI-RPA scheme. Also presented in Sec.~\ref{sec:results} are the fully converged atomic RPA correlation energies for the first 36 elements,
based on an iterative determination of the eigenspectrum of the $\chi^0 v$ operator. Section~\ref{sec:conclusion} concludes this paper. 
A detailed derivation of the frequency-dependent linear response theory, and the radial Sternheimer equation for spherical systems are
presented in Appendix~\ref{appendixA} and \ref{appendixb}, respectively. Appendix~\ref{appendixc} describes how the radial Sternheimer equation
is solved using the finite-difference method, and finally Appendix~\ref{appendixd} demonstrates how the RPA correlation energy converges
with respect to the spectrum of $\chi^0v$ and the highest angular momentum $L_{max}$.

\section{Theoretical Formulation}
\label{sec:theory}
In this section, the major equations to calculate the RI-RPA correlation 
energy are presented. The key quantity in RI-RPA calculations is the KS response function represented in terms of auxiliary basis functions (ABFs).
We illustrate how the response function matrix $\chi^0_{\mu\nu}$ is calculated using the Sternheimer approach, and this is compared to the conventional SOS scheme to evaluate $\chi^0_{\mu\nu}$. The essential difference between the two schemes lies in how the first-order wave function is computed.
Along the way, we analyze the source of the BSIE that plagues the SOS scheme and explain why it can be eliminated via the Sternheimer approach.

Derived from the adiabatic connection fluctuation-dissipation theorem \cite{Gunnarsson/Lundqvist:1976,Langreth/Perdew:1977}, the RPA correlation energy is given by \cite{Langreth/Perdew:1977,Gunnarsson/Lundqvist:1976}
\begin{equation}
 E_c^\mathrm{RPA}=\frac{1}{2\pi}\int_0^\infty {{\rm d}\omega}Tr[(\ln(1-\chi^0(i\omega)\nu)+\chi^0(i\omega)\nu)]
 \label{eq:EcRPA}
 \end{equation}
 where $\chi^0$ and $v$ represent the non-interacting density response function and the bare Coulomb interaction, respectively. 
 Physically, $\chi^0$ describes the density response of the KS system with respect to the change of
 the KS effective potential $v_{eff}(\bfr)$. In the real space and imaginary frequency domain, $\chi^0$ is formally given by  
 \begin{equation}
 \label{eq:define chi}
 \chi^0(\bm{r},\bm{r^\prime},i\omega)=\frac{\delta n(\bm{r},i\omega)}{\delta v_{eff}(\bm{r^\prime},i\omega)}
 \end {equation}
where $\delta v_{eff}(\bm{r^\prime},i\omega)$ represents the change of the effective potential on a spatial point $\bm{r^\prime}$ and at
imaginary frequency $\omega$, and $\delta n(\bm{r},i\omega)$  is the induced change of particle density on a spatial point $\bm{r}$.
Within the KS scheme, the electron density is expressed as a summation over all occupied states:
\begin{equation}
\label{eq:density}
n(\bm{r})=\sum_{i}^{occ}\psi_i(\bm{r})\psi_i^*(\bm{r})
\end {equation}
Combining Eq.~\ref{eq:density} and Eq.~\ref{eq:define chi}, one obtains
\begin{equation}
\label{eq:chi1}
 \chi^0(\bm{r},\bm{r^\prime},i\omega)=\sum_{i}^{occ}\frac{\delta \psi_i(\bm{r},i\omega)}{\delta v_{eff}(\bm{r^\prime},i\omega)}\psi_i^*(\bm{r})+ \mathrm{c.c.}
\end{equation}
where c.c. denotes the complex conjugate.
Within RI 
approximation, we use a set of ABFs $\{P_u(\bm{r})\}$ to represent the density response 
function, i.e.,
\begin{equation}
\label{eq:relation chi}
\chi^0(\bm{r},\bm{r^\prime},i\omega)=\sum_{\mu}\sum_{\nu}P_{\mu}(\bm{r})\chi_{\mu\nu}^0(i\omega)P_{\nu}^*(\bm{r^\prime})\, .
\end{equation}
For convenience, one can define its ``dual'' quantity $\tilde{\chi}_{\mu\nu}^0(i\omega)$ as
\begin{equation}
\label{eq:dual_chi0}
\tilde{\chi}_{\mu\nu}^0(i\omega)=\iint d\bfr d\bfrp P_{\mu}(\bm{r}) \chi^0(\bm{r},\bm{r^\prime},i\omega) P_{\nu}(\bm{r}')\, ,
\end{equation}
and obviously
\begin{equation}
    \tilde{\chi}_{\mu\nu}^0(i\omega) = \sum_{\mu'\nu'} S_{\mu,\mu'} \chi_{\mu'\nu'}^0(i\omega) S_{\nu'\nu}\, 
\end{equation}
where 
\begin{equation}
   S_{\mu,\mu'} = \int d\bfr  P_{\mu}(\bm{r}) P_{\mu'}(\bm{r})
\end{equation}
are the overlap integrals between the ABFs.
For monoatomic systems with which we are concerned in the present work, the ABFs are orthonormal by construction, i.e., $S_{\mu,\mu'} = \delta_{\mu,\mu'}$, and
hence $\tilde{\chi}_{\mu\nu}=\chi_{\mu\nu}$.

Combining Eq.~\ref{eq:dual_chi0} and Eq.~\ref{eq:chi1}, the matrix elements of the response function can be expressed as,
\begin{equation}
\label{eq:chi0munu}
\chi_{\mu\nu}^0(i\omega) = \tilde{\chi}_{\mu\nu}^0(i\omega) = \sum_i^{occ}\int P_{\mu}^*(\bm{r})\psi_i^*(\bm{r})\psi_{i\nu}^{(1)}(\bm{r},i\omega) d\bm{r}+ \rm c.c.
\end{equation}
where
\begin{equation}
\psi_{i\nu}^{(1)}(\bm{r},i\omega)=\int \frac{\delta \psi_i(\bm{r},i\omega)}{\delta v_{eff}(\bm{r^\prime},i\omega)}P_{\nu}(\bm{r^\prime}) d{\bm{r^\prime}}\, .
\label{eq:first_order_psi}
\end{equation}
can be interpreted as the first-order change of the wavefunction $\psi_i(\bfr)$ induced by  an ``external 
purturbation'' $P_{\nu}(\bm{r^\prime})$ to the KS system. In another word, the matrix element $\tilde{\chi}_{\mu\nu}^0(i\omega)$ is
nothing but the projection of the first-order density change of the KS system as induced by an ``potential'' $P_\nu(\bfrp)$ 
to a function $P_\mu(\bfr)$.  

With the above understanding, the task of evaluating the non-interacting response function matrix $\chi_{\mu\nu}^0(i\omega)$
boils down to the evaluation of the frequency-dependent first-order wave function $\psi_{i\nu}^{(1)}(\bm{r},i\omega)$ 
as given by Eq.~\ref{eq:first_order_psi}. 
In fact, it is the different ways of evaluating $\psi_{i\nu}^{(1)}(\bm{r},i\omega)$ that distinguish
the conventional SOS scheme and the Sternheimer approach.
To illustrate this point, we start with the single-particle KS Hamiltonian of the system $H^{(0)}$, which in general has the following form,
\begin{equation}
    H^{(0)}=-\frac{1}{2}\nabla^2_\bfr+v_{eff}(\bm{r})
\end{equation}
where the effective potential $v_{eff}(\bm{r})$ is composed of three terms,
\begin{equation}
    v_{eff}(\bm{r})=v_{ext}(\bm{r})+v_{H}(\bm{r})+v_{xc}(\bm{r})\, .
\end{equation}
with $v_{ext}(\bm{r})$, $v_{H}(\bm{r})$, $v_{xc}(\bm{r})$ representing the external potential, Hartree potential, and 
exchange-correlation (XC) potential, respectively.
Let's now consider adding a small frequency-dependent perturbation $V^{(1)}(\bm{r})e^{i\omega t}$ to the Hamiltonian $H^{(0)}$. 
According to the frequency-dependent linear 
response theory as described in Ref.~\cite{marques2012fundamentals} and in Appendix~\ref{appendixA}, 
the linear response of the system to this perturbation is governed by the following frequency-dependent Sternheimer equation,
\begin{equation}
    (H^{(0)}-\epsilon_i+i\omega)\psi_{i}^{(1)}(\bm{r},i\omega)=(\epsilon_i^{(1)}-V^{(1)})\psi_i(\bm{r})
    \label{eq:st}
\end{equation}
where $\psi_i(\bm{r})$ and $\epsilon_i$ are KS orbitals and their orbital energies, and
$\psi_{i}^{(1)}(\bm{r},i\omega)$ and $\epsilon_i^{(1)}$ are the first-order changes 
of the orbitals and orbital energies. The question is how to solve this equation to obtain $\psi_{i}^{(1)}(\bm{r},i\omega)$.

\subsection{The SOS approach}
To solve Eq.~\ref{eq:st}, one straightforward way is to expand $\psi_{i}^{(1)}(\bm{r},i\omega)$ in terms of $\{\psi_i(\bm{r})\}$, i.e.,
the entire spectrum of $H^{(0)}$. This leads to the renowned expression of the first-order perturbation theory,
\begin{equation}
   \psi_{i}^{(1)}(\bm{r},i\omega)=\sum_{j\neq{i}}\frac{\int{\rm d}\bm{r^\prime}\psi_j^*(\bm{r^\prime})V^{(1)}(\bm{r^\prime})\psi_i(\bm{r^\prime})}{\epsilon_i-\epsilon_j-i\omega}\psi_j(\bm{r})
\label{eq:one order wavefunction}
\end{equation}
where the summation of $j$ goes over both occupied (except for the state $i$) and unoccupied states, and
\begin{equation}
  \epsilon_i^{(1)}=\left \langle \psi_i\right |V^{(1)} \left | \psi_i \right \rangle \, . 
\label{eq:1 order energy 3 dimension}
\end{equation}

Plugging Eq.~\ref{eq:one order wavefunction} into Eq.~\ref{eq:chi0munu}, one arrives at the widely used ``SOS'' expression of the response function matrix,
\begin{eqnarray}
    \chi_{\mu\nu}^0(i\omega) & = &\int\int{\rm d}\bm{r}{\rm d}\bm{r^\prime}\sum_{i}^{occ}\sum_{j\neq{i}}\frac{P_\mu^*(\bm{r})\psi_i^*(\bm{r})\psi_j^*(\bm{r^\prime})\psi_j(\bm{r})\psi_i(\bm{r^\prime})P_\nu(\bm{r^\prime})}{\epsilon_i-\epsilon_j-i\omega}+ c.c.  \label{eq:auxil_space_chi0} \\
    & = &\int\int{\rm d}\bm{r}{\rm d}\bm{r^\prime}\sum_{i}^{occ}\sum_{j}^{unocc}\frac{P_\mu^*(\bm{r})\psi_i^*(\bm{r})\psi_j^*(\bm{r^\prime})\psi_j(\bm{r})\psi_i(\bm{r^\prime})P_\nu(\bm{r^\prime})}{\epsilon_i-\epsilon_j-i\omega}+ c.c.
    \label{eq:auxil_space_chi0_1}
\end{eqnarray}
From Eq.~\ref{eq:auxil_space_chi0} to Eq.~\ref{eq:auxil_space_chi0_1}, we have used the fact that for occupied $j$, swapping the the indices $i$
and $j$ in Eq.~\ref{eq:auxil_space_chi0} yields opposite contributions which cancel out each other. Hence the summation over occupied states $j$ doesn't
contribute to the response function. 

Obviously, the computation of $\chi_{\mu\nu}^0$ in Eq.~(\ref{eq:auxil_space_chi0}) contains two summations: One is the summation over all occupied 
states, coming from the 
charge density calculation; the other is the summation over all unoccupied ones, which comes from 
the first-order perturbation expansion.
Within a local atomic-orbital framework, the actual number and quality of the calculated unoccupied states are determined by the employed basis set. 
In practical calculations, however, the basis set is always finite in size, and the sum of unoccupied states is always limited, resulting in the BSIE.

With the response function matrix $\chi_{\mu\nu}^0$ obtained, one stills needs to compute the Coulomb matrix, i.e., the matrix representation
of the Coulomb operator within the auxiliary basis set are given by \cite{Ren/etal:2012b},
\begin{equation}
    V_{\mu\nu}=\int\int{\rm d}\bm{r}{\rm d}\bm{r^\prime}\frac{P_\mu^*(\bm{r})P_\nu(\bm{r^\prime})}{\lvert {\bm{r}-\bm{r^\prime}}\rvert} \, .
    \label{eq:auxil space Coulomb}
\end{equation}
In standard RPA implementation in FHI-aims, the SOS scheme is employed, whereby 
Eqs.~(\ref{eq:EcRPA}), (\ref{eq:auxil_space_chi0}), and (\ref{eq:auxil space Coulomb}) are combined to yield the RPA correlation energies.

\subsection{The Sternheimer approach}
\label{method:Sternheimer}
As mentioned above, the BSIE inherent to the SOS scheme stems from the summation over unoccupied states, 
which is limited by the finite AO basis set to expand the eigenstates of $H_0$. Alternatively, one may solve the Sternheimer equation (Eq.~\ref{eq:st}) on a dense real-space grid by numerical integration to obtain accurate first-order wave function that is not limited by
the original AO basis set. Then, using Eq.~\ref{eq:chi0munu}, one can obtain a density-response function matrix 
$\chi^0_{\mu\nu}$ that is essentially
free of SPBS error.  Further using  Eqs.~\ref{eq:EcRPA} and \ref{eq:auxil space Coulomb}, the RPA correlation energy that
is converged with respect to SPBS can be obtained. Of course, the RPA correlation energy calculated in this way still
suffers from incompleteness error of the ABS. However, the ABS error is controllable and can usually be made much smaller than the BSIE of the AOs. Later in this work, we will also discuss how to eliminate the second source of errors. 
For general molecular or solid systems, accurately solving the Sternheimer equation on a dense, three-dimensional real-space grid, 
can become prohibitively expensive. However, for an atomic system with spherical symmetry, one can separate variables in the spherical coordinate system.
Consequently, the three-dimensional Sternheimer equation reduces to an one-dimensional radial differential equation which can then be 
solved arbitrarily accurately.
Considering an isolated atom with spherical symmetry, the effective potential in its KS Hamiltonian only depends on the distance, i.e.,
\begin{equation}
    H^{(0)}(\bm{r})=-\frac{1}{2}\bigtriangledown^2+V_{eff}(r)\, ,
    \label{eq:H^0}
\end{equation}
 and the KS eigenstates are given by a radial function multiplied by spherical harmonics\, ,
 \begin{equation}
\psi_i(\bm{r})=u_{i,l}(r)Y_l^m(\theta,\phi)
\label{eq:psi^0}
\end{equation}
In FHI-aims, the ABFs also have the form of a radial function multiplied by a spherical harmonic function,
\begin{equation}
P_\mu(\bm{r})=\frac{\xi_\mu(r)}{r}Y_L^M(\theta,\phi) \, .
\end{equation}
As discussed above, the response function matrix in RI-RPA can be obtained by taking the ABFs themselves as the
perturbation to the KS Hamiltonian,
\begin{equation}
V_\mu^{(1)}(\bm{r})=V_\mu^{(1)}(r)Y_L^M(\theta,\phi)=P_\mu(\bm{r})
\label{eq:auxil_P}
\end{equation}
which means that radial part of the perturbative potential $V_\mu^{(1)}(r)=\xi_\mu(r)/r$.

It should be noted that we use real spherical harmonic functions in this paper which are the linear combination of complex spherical harmonics. In the presence of degeneracy, we need to diagonalize $V_\mu^{(1)}(\bm{r})$ in the degenerate subspace, and so it is convenient to ensure the hermiticity of $ V_\mu^{(1)}(\bm{r})$ . 
In FHI-aims, the real spherical harmonic function has the following expression:
\begin{equation}
Y_l^m(\theta,\phi)	=\left\{
		\begin{aligned}
		\quad&\sqrt{2}Re y_l^m(\theta,\phi),\quad  &m > 0 \\ 
		\quad&y_l^m(\theta,\phi)  ,\quad &m = 0 \\
		\quad&-\sqrt{2}Im y_l^m(\theta,\phi),\quad  &m < 0
	\end{aligned}
	\right.
\end{equation}
where, $y_l^m(\theta,\phi)$ denotes the original complex spherical harmonic function.


For a spherically symmetric system, if it is perturbed by a multipole potential, i.e., a potential given by a radial functional multiplied by
spherical harmonics, then the 1st-order wave function is the superposition of a series of radial functions multiplied by 
spherical harmonics of varying angular momentum channels,
\begin{equation}
\psi_{i\mu}^{(1)}(\bm{r},i\omega)=\sum_{l^\prime m^\prime}u_{i\mu,l^\prime m^\prime}^{(1)} (r,i\omega)Y_{l^\prime}^{m^\prime}(\theta, \phi)
\label{eq:psi_1_comp}
\end{equation}
Notice that here we use $\left \{ {l,m} \right \} $ for the angular momentum of zeorth-order KS eigenstates,  $\left \{ {l^\prime,m^\prime} \right \} $ 
for the different angular momentum component of the first-order wavefunctions, and $\left \{ {L,M} \right \} $ for the angular momentum of the perturbation (here the ABFs).

Combining Eqs.~\ref{eq:st},  \ref{eq:psi^0}, \ref{eq:auxil_P}, and \ref{eq:psi_1_comp}, we can obtain, by separating variables, the one-dimensional differential equation satisfied by the radial component of the first-order wavefunction,
\begin{equation}
\left(-\frac{1}{2r}\frac{\partial^2}{\partial r^2}r+\frac{l^\prime(l^\prime+1)}{2r^2}+V_{eff}(r)-\epsilon_i+i\omega\right)u_{i\mu,l^\prime m^\prime}^{(1)}(r,i\omega)=G_{l^\prime Ll}^{m^\prime Mm}\left(\epsilon_{i-rad}^{(1)}\delta_{ll^\prime}\delta_{mm^\prime}-V_\mu^{(1)}(r)\right)u_{i,l}(r)\, .
\label{eq:radial st}
\end{equation}
The details of the derivation are given in Appendix~\ref{appendixb}.
In Eq.~\ref{eq:radial st}, $G_{l^\prime Ll}^{m^\prime Mm}$ are the real Gaunt coefficients, which represent the angular integrals of triple real spherical harmonic functions,
\begin{equation}
G_{l^\prime Ll}^{m^\prime Mm}=\int_{0}^{2\pi}{\rm d} {\phi}\int_{0}^{\pi}sin(\theta){\rm d} {\theta}{Y_{l^\prime}^{m^\prime}}(\theta,\phi)Y_{L}^{M}(\theta,\phi)Y_{l}^{m}(\theta,\phi)
\end{equation}
Furthermore, different from $\epsilon_i^{(1)}$ in Eq.\ref{eq:st}, $\epsilon_{i-rad}^{(1)}$ in Eq.~\ref{eq:radial st} is 
the radial integral of the 1th-order energy,
\begin{equation}
\epsilon_{i-rad}^{(1)}=\int{\rm d} {r}u_{i,l}^*(r)V_\mu^{(1)}(r))u_{i,l}(r)
\end{equation}
According to the symmetry properties of spherical harmonic functions, real Gaunt coefficients are zero in many cases, yielding the ``selection rule" in 
optical excitation process. For further details, please refer to Ref.~\citenum{talman2011multipole}.  
 When the Gaunt coefficient equals zero, the right-hand side of Eq.~\ref{eq:radial st} is zero, meaning that the radial Sternheimer equation
 has no non-trivial solutions. Therefore, due to the constraints imposed by the Gaunt coefficients, we only need to use Eq.~\ref{eq:radial st} to calculate $u_{i\mu,l^\prime m^\prime}^{(1)}(r,i\omega)$ for those
 $\left \{ {l^\prime,m^\prime} \right \}$ whereby the Gaunt coefficients are nonzero. 

One further important problem to note is that the derivation of Eq.~{\ref{eq:radial st} } is based on non-degenerate perturbation theory. When the angular momentum $l$ of the zeroth-order wave function is greater than 0, the energy level of the ground state has a $(2l+1)$-dimensional degeneracy. According to the degenerate perturbation theory, if the perturbation Hamiltonian is diagonal in this subspace of degenerate eigen-orbitals, we can still use the same formalism of the non-degenerate perturbation theory. Otherwise, we need to linearly combine the degenerate
zeroth-order wavefunctions to get ``good'' zeroth-order orbitals. The combination coefficients can be obtained by diagonalizing the matrix block
of the perturbed Hamiltonian within the degenerate subspace, yielding ``good'' zeroth-order wavefunctions  
as
\begin{equation}
\psi_i(\bm{r})=u_{i,l}(r)\sum_{k=-m}^{k=m}c_{i,k} Y_l^k(\theta,\phi)\ ,
\label{eq:psi^0 degenrate}
\end{equation}
 
 With the above ``good'' zeroth-order eigen-orbitals $\psi_i$ (or equivalently the combination coefficients $c_{i,k}$), the one-dimensional radial Sternheimer equation in the degenerate case can be obtained similarly by separating the variables of Eq.~\ref{eq:st}:
 \begin{equation}
(-\frac{1}{2r}\frac{\partial^2}{\partial r^2}r+\frac{l^\prime(l^\prime+1)}{2r^2}+V_{eff}(r)-\epsilon_i+i\omega)u_{i\mu,l^\prime m^\prime}^{(1)}(r,i\omega)=\sum_{k=-m}^{k=m}c_{i,k}(\epsilon_{i}^{(1)}\delta_{ll^\prime}\delta_{km^\prime}-G_{l^\prime Ll}^{m^\prime Mk}V_\mu^{(1)}(r))u_{i,l}(r) \, .
\label{eq:radial st degenrate}
\end{equation}
 As in the nondegenerate case, we only need to use Eq.~\ref{eq:radial st degenrate} to calculate $u_{i\mu,l^\prime m^\prime}^{(1)}(r,i\omega)$ corresponding to
 $\left \{ {l^\prime,m^\prime} \right \}$ whereby the right-hand side of the above equation is nonzero.
 
\section{Implementation details} 
\label{sec:imple}
\begin{figure}[htbp]
    
    \centering
    
    \subfloat{\includegraphics[scale=0.15]{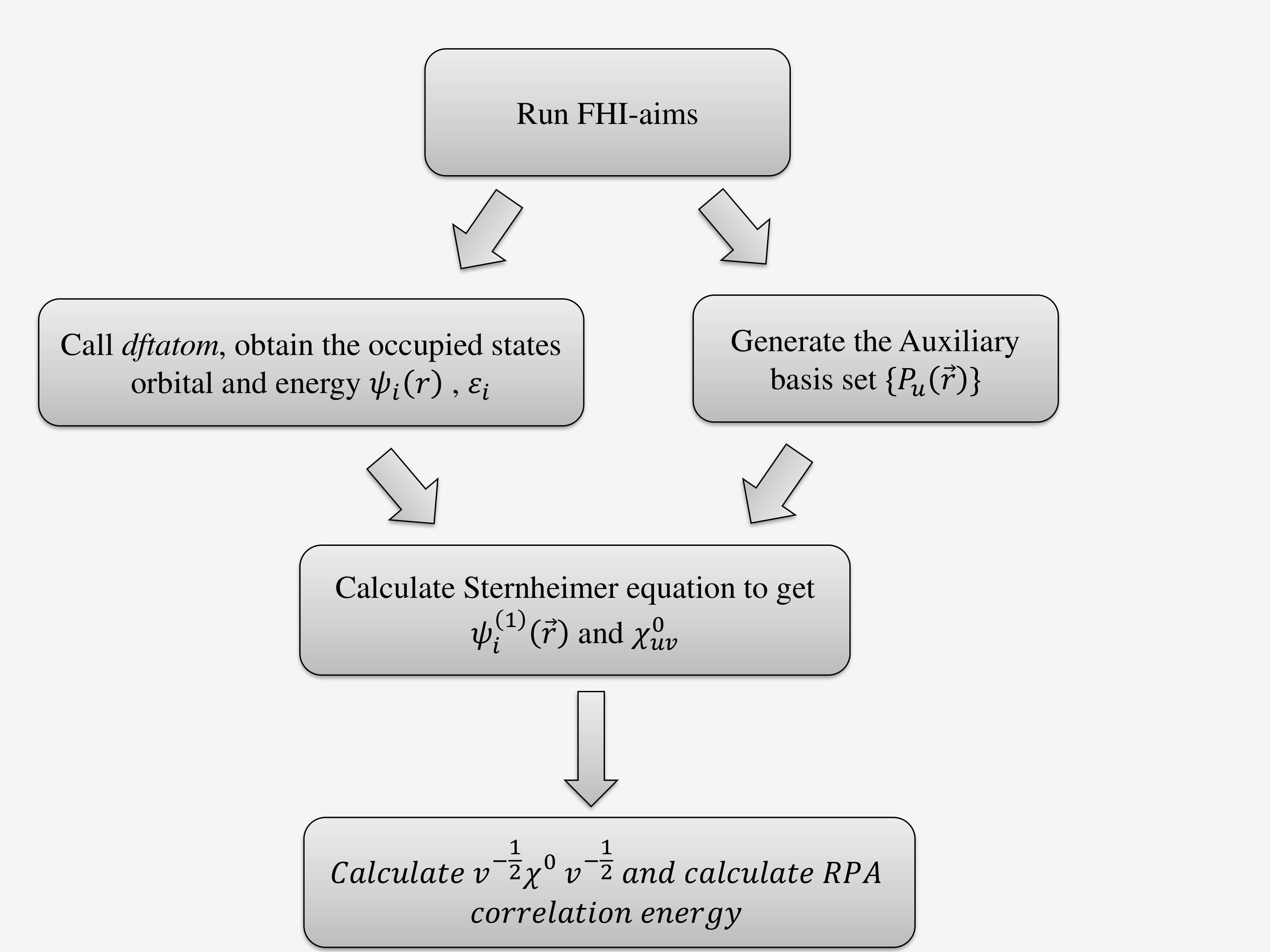}}
    \caption{\label{fig:flowchart} Major steps for computing the RI-RPA correlation energy using the Sternheimer approach.}
    
\end{figure}

The RI-RPA based on the SOS formalism had been implemented in FHI-aims previously \cite{ren2012resolution}. 
In the present work, the Sternheimer approach to compute the RI-RPA correlation energy, as described in Sec.~\ref{method:Sternheimer}
is implemented in FHI-aims, interfaced with the DFT atomic solver ``dftatom'' package \cite{vcertik2013dftatom}. 
Specifically, FHI-aims calls dftatom to solve self-consistently the KS-DFT equation of a single atom,
obtaining the spherically symmetric effective potential
$V_{eff}(r)$, and the occupied KS orbital energies $\epsilon_i$ and radial functions $u_{il}(r)$ for
the unperturbed Hamiltonian. In the meantime, FHI-aims also generates the ABFs for a given set of single-particle AO basis functions
using the automatic procedure as described in Ref.~\cite{ren2012resolution}.
With these inputs, we solve the radial Sternheimer equation on a dense grid to obtain the first-order wave function, and eventually
the non-interacting KS density response function represented within a set of ABFs. Finally, the computation of RI-RPA correlation energy 
proceeds as usual \cite{ren2012resolution} with the above computed $\chi^0_{\mu\nu}$. The entire procedure is illustrated
by the flow diagram presented in Fig.~\ref{fig:flowchart}.

To solve the radial Sternheimer equation, we use the finite difference approximation to transform the second-order differential equation into a linear simultaneous equation of the form $AX=B$, which is then solved using Thomas algorithm \cite{press2007numerical}. 
 This method is suitable for solving complex second-order differential equations \cite{betzinger2015precise}. 
 Further details are provided in Appendix~\ref{appendixc}.
 
 In the present work, the LDA is used to generate the zeroth-order Hamiltonian and wavefunctions
 employed in the Sternheimer equation, although the generalization to references based on other exchange-correlation functionals is straightforward. 
 For the RI approximation, the Coulomb-metric RI-V flavor 
 \cite{ren2012resolution} is used in this work.
 As for the imaginary frequency grid,  a modified Gauss-Legendre grid with 100 
 points is used for the frequency integration in Eq.~\ref{eq:EcRPA}. Alternatively, the more efficient minimax grid
 can also be used \cite{Klimes/Kaltak/Kresse:2014}. In both cases, a sub-meV convergence of the RPA correlation energy
 with respect to the frequency grid can be readily achieved.

 \section{Results}
 \label{sec:results}
 
 \subsection{Converged RPA correlation energies for atoms}
The basis-set dependence of the RI-RPA correlation energy obtained via the SOS scheme is 
reflected in two aspects: the size of the SPBS and that of ABS. The first
determines the number of unoccupied states, which in turn determines the precision
of the response function $\chi^0(\bfr,\bfrp,i\omega)$ in real space; the second determines
the rank of the response function matrix, which governs how well $\chi^0(\bfr,\bfrp,i\omega)$ is represented
in a matrix form. Naturally,
the completeness of the ABS also affects the calculated results, which is usually referred to as the RI errors.
In FHI-aims, the ABFs are generated according to the automatic procedure as described in Ref.~\citenum{ren2012resolution}  
and  Ref.~\citenum{ihrig2015accurate}. Within this procedure, the number of ABFs are controlled by two
parameters: the highest angular momentum of the ABFs, $max\_l\_{aux}$, and a threshold $\eta_{orth}$ 
which is used in the Gram-Schmidt orthonormalization procedure to remove the redundancy of the  ``on-site'' pair products of radial functions of AOs.
In the tests below, we chose $max\_l\_{aux}=6$, and $\eta_{orth}= 10^{-4}$, which works rather well
for the SOS scheme, i.e., the RI error is vanishing small. 
We test two types of basis sets: NAO-VCC-nZ \cite{IgorZhang/etal:2013} for NAOs and aug-cc-pVXZ for 
\cite{Dunning:1989,yousaf2009optimized} for GTOs, and check their convergence behavior for computing
the RI-RPA correlation energy in both the SOS and Sternheimer schemes.

For the reference numbers adopted in this work, the ``hard-wall cavity'' approach as developed in  Ref.~\cite{Jiang/Engel:2005} is used, which has been demonstrated to yield highly converged RPA correlation energies
for closed-shell atoms \cite{jiang2007random}. 
Within this approach, the atom is placed in a cavity subject to a hard-wall confining potential imposed at
a finite but large radius $R_\text{max}$. The radial KS equation is then solved on an one-dimensional dense grid mesh, yielding series of eigenstates characterized by principal quantum number $n$ and angular momentum quantum number $l$.  These states 
consist in a systematic representation of the unoccupied space. Indeed, employing the thus obtained dense eigenstates
in the SOS scheme, one can attain highly converged RPA correlation energies for atoms \cite{jiang2007random}.
In addition to $R_\text{max}$, other convergence parameters in this approach are $n_{max}$ and $l_{max}$ 
denoting the highest principal and angular momentum quantum numbers, respectively.
Note that $n_{max}$ is typically of the order of a few hundreds, to achieve sub-mHa convergence in all-electron RPA correlation energy for each 
angular momentum $l$. Thus the total number of unoccupied states in this approach is huge and it is difficult to extend it
to molecular systems. In Ref.~\cite{jiang2007random}, the reported RPA correlation energies are calculated on top of the KS
reference of the optimized effective potential (OEP) at the exact exchange level (denoted as RPA@OEPx). To facilitate the comparison 
to the RPA results in this work, the ``hard-wall cavity'' approach is used here on top of the LDA
starting point.

\begin{figure}[htbp]
    
    \centering
    \subfloat[\label{fig:st_NAO}]{\includegraphics[scale=0.4]{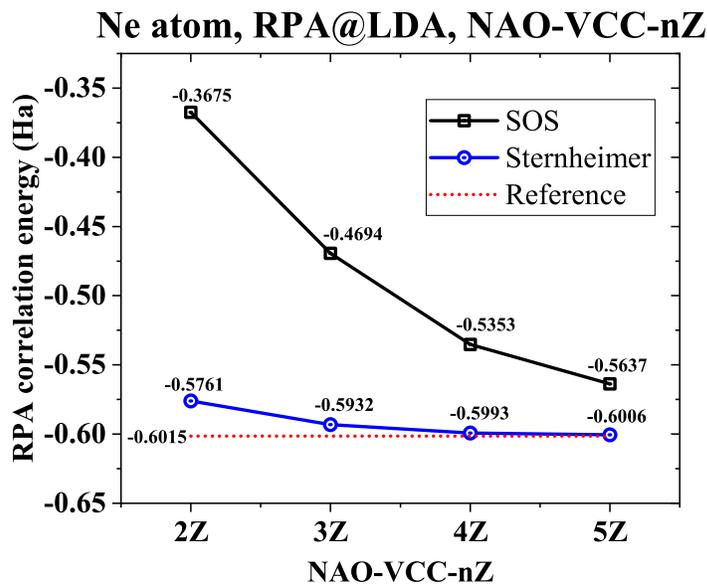}}
    \\
    \subfloat[\label{fig:st_aug}]{\includegraphics[scale=0.4]{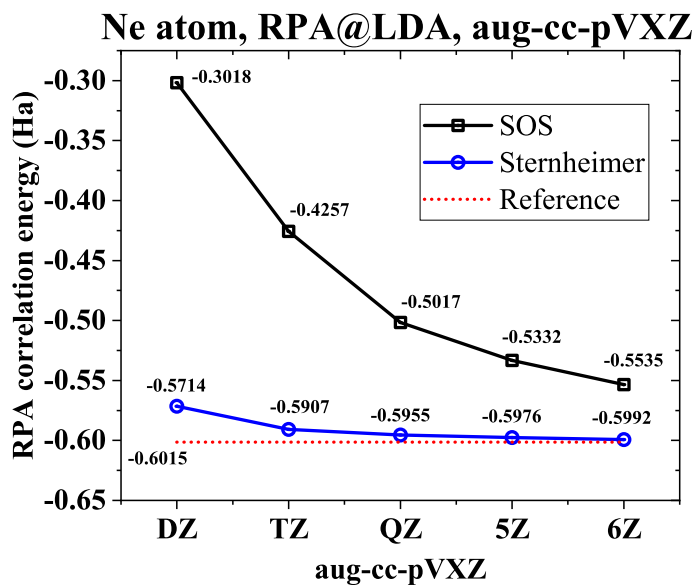}}
    \caption{The RI-RPA correlation energy for the Ne atom as a function of the basis set size as obtained from the SOS and Sternheimer schemes. Results in panel (a) and (b) are calculated using NAO (NAO-VCC-nZ) and GTO (aug-cc-pVXZ) basis sets, respectively. The red dash line marks the reference result obtained using the ``hard-wall cavity'' approach \cite{jiang2007random}.}
    \label{fig:Sternheimer-vs-SOS}
\end{figure}

\begin{figure}[htbp]
    \centering
    \subfloat[\label{fig:NAO BASIS ERROR}]{\includegraphics[scale=0.4]{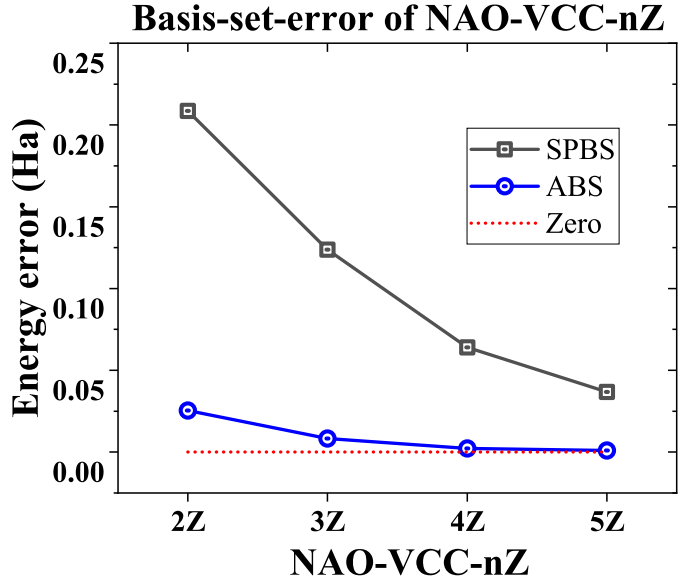}}
    \\
    \subfloat[\label{fig:AUG BASIS ERROR}]{\includegraphics[scale=0.4]{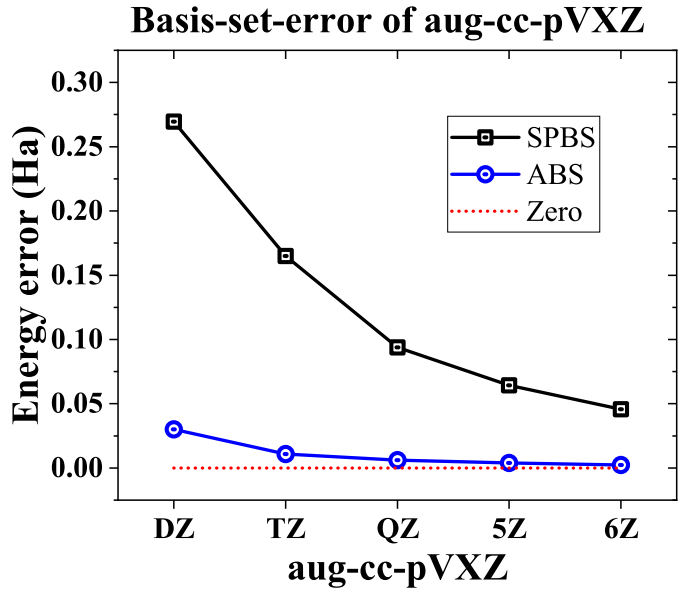}}
    \caption{Two types of basis set incompleteness errors as a function of the basis size for both NAOs (panel a) and GTOs (panel b). Black squares indicate the difference between the
    RPA correlation energies for the Ne atom as obtained from the SOS and Stermheimer schemes ($E_c^{RPA-SOS}(n)-E_c^{RPA-Sternheimer}(n)$); the
    blue circles indicate the difference between the energies obtained from Sternheimer scheme and the reference value 
    ($E_c^{RPA-Sternheimer}(n)-E_c^{RPA-ref}$).
    }
    \label{fig:SPBS VS ABS}
\end{figure}

In Fig.~\ref{fig:Sternheimer-vs-SOS}, the RPA 
correlation energy of the Ne atom are presented as a function of the basis set size for both the SOS and Sternheimer schemes. 
We emphasize again that, in the Sternheimer scheme, one only uses the ABS generated using the SPBS labelled in the $x$ axis, 
rather than using the SPBS to expand the KS orbitals. Panel (a) and (b) of Fig.~\ref{fig:Sternheimer-vs-SOS} present
the convergence behavior for NAOs (NAO-VCC-$n$Z with $n=2$-5) and GTOs (aug-cc-pVXZ with X$=$D, T, Q, 5, 6), respectively. 
The reference value here is computed
using the above-mentioned ``hard-wall cavity" approach with $l_{max}=6$ (to be consistent with $max\_l\_{aux}=6$ for the ABS), $n_{max}=300$ and 
$R_{max}=10$ Bohr.
From Fig.~\ref{fig:Sternheimer-vs-SOS}, one can clearly see that the RPA correlation energies based on the Sternheimer equation 
are much lower than their counterparts yielded by the SOS scheme, and can quickly converge to the reference value (up to 1 mHa) 
marked by horizontal dashed line, 
when the basis set increases from 2Z to 5Z for NAOs (Panel a). Similar convergence behavior is observed for GTOs from DZ to 6Z (Panel b). 
Remarkably, the RPA correlation energy obtained from the Sternheimer scheme using the 2Z-generated ABS is even lower than those obtained
from the conventional SOS scheme using the largest NAO-VCC-5Z or aug-cc-pV6Z basis sets. 
It should be understood that the difference between the RPA correlation energies given by the two schemes reflects the incompleteness error of the
SPBS, while the difference between the results based on the Sternheimer equation and the reference values reflects the incompleteness error of
the ABS. In Fig.~\ref{fig:SPBS VS ABS}, these two types of BSIEs are further compared.  It can be seen that, with the increase of the basis set size, the errors incurred by the SPBS and the ABS are both decreasing, but the error from the incompleteness of 
SPBS is dominating, and does not converge to zero even with the largest available NAO or GTO basis sets. As clearly shown in this session,
invoking the Sternheimer approach to compute numerically accurate first-order wave functions can eliminate the major source of BSIEs
in RI-RPA calculations.
%

Next, we look at the RPA correlation energies as obtained by the SOS and Sternheimer schemes for other elements
with atomic numbers from 2 to 18, 
and compare the obtained results to the reference values as obtained using the ``hard-wall cavity" approach \cite{jiang2007random}. 
\begin{figure}
	\centering
	\subfloat[\label{fig:NAO_he-ne}]{
	 \includegraphics[scale=0.4]{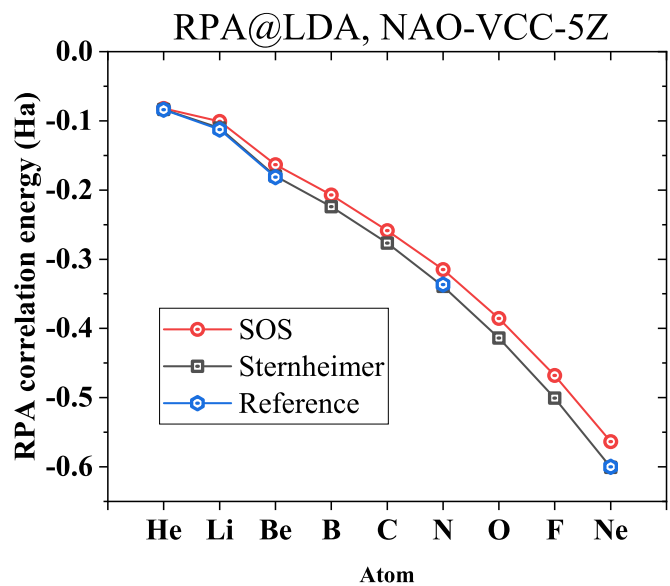}
                      	}%
                       \\
\subfloat[\label{fig:NAO_na-ar}]{
	 \includegraphics[scale=0.4]{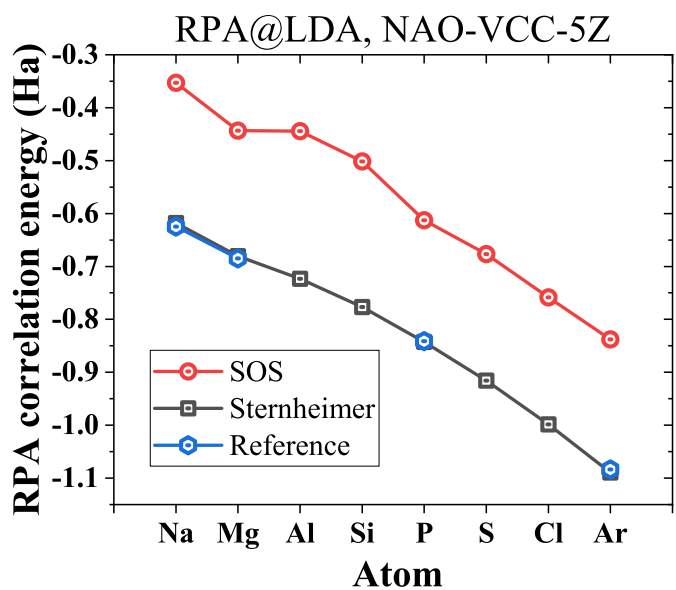}
                      	}
	\caption{RPA correlation energies for atoms from He to Ar. The SOS results are obtained 
                 using the NAO-VCC-5Z basis set and the Sternheimer results are obtained using the ABS generated from NAO-VCC-5Z.}
	\label{Fig:RPA_Ec}
\end{figure}
As is shown in Fig.~\ref{Fig:RPA_Ec}, for all atoms, the RPA correlation energies obtained from the Sternheimer scheme are very close to the
corresponding reference results, when they are available. In contrast, similar to the Ne atom case, the RPA correlation energies obtained using
the conventional SOS scheme are appreciably above the reference results, except for He. This trend gets more pronounced for heavier elements 
in the third row of the periodic table (Panel b). 
The series of results presented in Fig.~\ref{Fig:RPA_Ec} confirm the efficacy of the Sternheimer scheme.  



%

\subsection{Further analysis: First-order density in real space}

The significant difference between the RPA correlation energies yielded by the SOS and Sternheimer schemes
is due to their different underlying density response function matrices $\chi^0_{\mu,\nu}$, which can 
be further traced back to the difference in the first-order density $n^{(1)}(\bfr)$ as obtained
by the two schemes -- denoted as $\Delta n^{(1)}(\bfr)$ below. It will be instructive to visualize $\Delta n^{(1)}(\bfr)$, which will then
signify what is missing in the existing AO basis sets to accurately represent $n^{(1)}(\bfr)$.
This will hopefully provide us a guidance for further improving the existing basis sets for correlated calculations. 

Naturally, the shape of $n^{(1)}(\bfr)$, and hence that of $\Delta n^{(1)}(\bfr)$ depend on the actual applied perturbation. We found that for a spherically symmetric atomic system, the angular distributions of the first-order density and the perturbed Hamiltonian are always the same. 
Therefore, we only need to pay attention to the radial distribution of the first-order density difference.

\begin{figure}[htbp]
    \centering
    \subfloat{\includegraphics[scale=0.35]{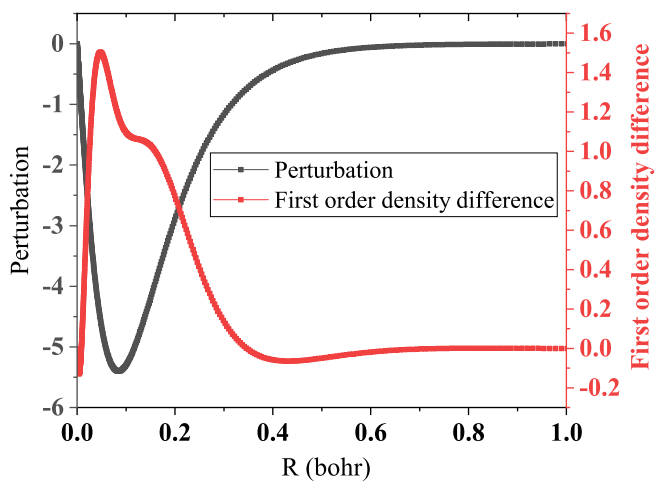}}
    \\
    \subfloat{\includegraphics[scale=0.35]{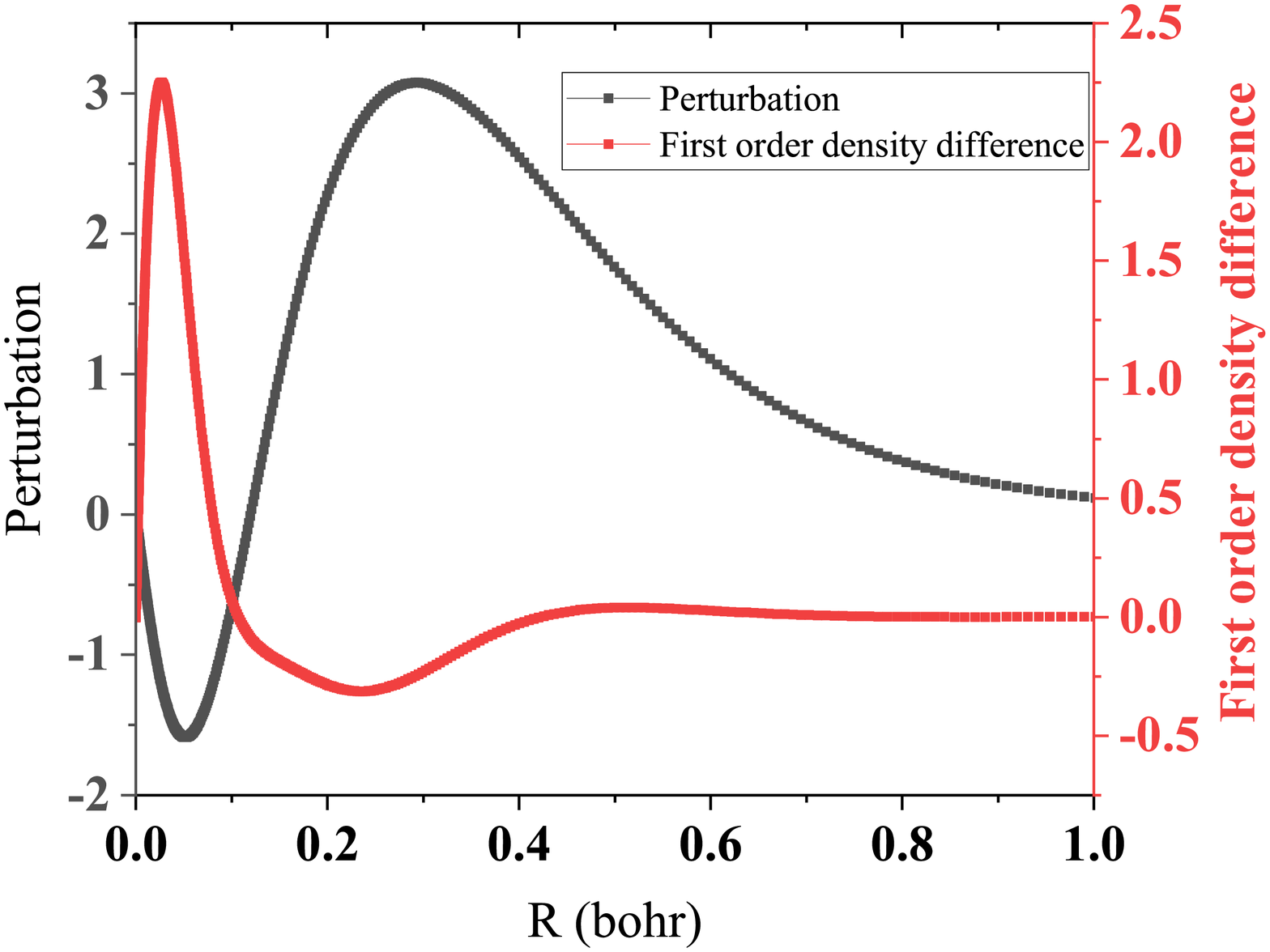}}
    \\
    \subfloat{\includegraphics[scale=0.35]{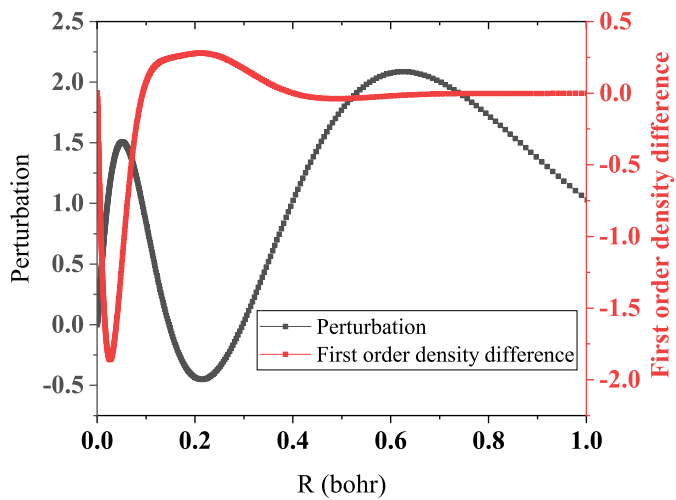}}
    \caption{First order density differences $\Delta n^{(1)}(\bfr)$ of Ar atom under $p$-type perturbations with different radial shapes
    .
    }
    \label{fig:1order_density_diff}
\end{figure}

As a test example, we add a $Y^0_1-type$ (i.e. $p_z$-type) perturbative potential to the Ar atom, and evaluate the difference in 
$n^{(1)}(\bfr)$ yielded by the two schemes. 
Here, the SPBS used in the SOS scheme is NAO-VCC-5Z, a rather large basis set
in practical calculations.
The radial part of 
$\Delta n^{(1)}(\bfr)$ and that of the applied perturbative potential are plotted in Fig.~\ref{fig:1order_density_diff}. In the upper, middle, and lower panels of Fig.~\ref{fig:1order_density_diff},  three different shapes of the 
perturbative potentials (different $p$-type radial ABFs), together with the resultant radial distributions of $\Delta n^{(1)}(\bfr)$ 
are plotted.
It can be seen that, although the detailed behavior of $\Delta n^{(1)}(\bfr)$ depends on the applied disturbance, there are some common features:
First, the error in the first-order density yielded by the SOS method roughly occurs in the region between 0  to 0.5 Bohr around the nucleus.
Second, the radial distribution of $\Delta n^{(1)}(\bfr)$ is similar under different perturbations, with a peak near the nucleus.
We also tested other ($s$, $d$, or $f$) types of perturbations, and the resultant $\Delta n^{(1)}(\bfr)$ show similar behavior as that illustrated
in Fig.~\ref{fig:1order_density_diff}. The insights gained from the behavior of $\Delta n^{(1)}(\bfr)$ is
instructive for developing complementary AOs to mitigate the BSIE for molecular and condensed materials. Research along this direction goes beyond the
scope of the present paper and will be explored in the future work. 

\subsection{Optimization of ABFs}
\label{sec:ABF-opt}
As demonstrated above, 
The RI-RPA correlation energy obtained using the Sternheimer scheme is free of SPBS error, but still depends on the quality of the employed ABS to
represent the non-interacting response function $\chi^0$. 
In Ref.~\citenum{eshuis2010fast}, it is pointed out that the Coulomb-metric RI-RPA 
correlation energy is variational, meaning that it is lower-bounded by the true
RPA correlation energy that is free of the RI errors.  
In other words, by systematically increasing the variational space of the ABFs, 
the RI error of RI approximation decreases, and the RI-RPA correlation energy should in principle gradually converge 
to the true energy that is free of any BSIE. 

Here we show that the variational property of RI-RPA can be exploited to generate optimized ABFs,
picked up from a large pool of candidate functions.
The candidate ABFs adopted here still have the form of a radial function multiplied by 
spherical harmonics, but for simplicity are restricted to hydrogen-like functions.
The shape of hydrogen-like functions is controlled by the principal and angular quantum numbers, and by 
the effective charge $Z$ of the nucleus, i.e.,
\begin{equation}
\label{eq:hy-like}
R_{n l}(r)=\sqrt{\left(\frac{2 Z}{n a}\right)^{3} \frac{(n-l-1) !}{2 n(n+l) !}}\left(\frac{2 Z r}{n a}\right)^{l} L_{n-l-1}^{2 l+1}\left(\frac{2 Z r}{n a}\right) \mathrm{e}^{-Z r /(n a)} \, ,
\end{equation}
where $L_{n-l-1}^{2 l+1}$ in upper equation represents the associated Laguerre polynomial.
The optimization procedure adopted here follows that used in Ref.~\citenum{IgorZhang/etal:2013}.
Namely, during the optimization process, we add trial hydrogen-like functions from a large pool of candidate functions one by one,
and each time observe how the obtained RI-RPA correlation energy change with the effective 
charge $Z$. The one that yields the largest drop of the obtained energy will be picked up and added to the
existing ABS, until a pre-set accuracy is reached. 

Due to the orthogonality of the spherical harmonic functions, the ABFs with different angular momenta are independent of each other, and
hence we optimize the ABFs belonging to different angular momentum channels separately.
The convergence criterion of ABF optimization for each angular momentum 
is set to be $10^{-4}$ Ha, under which the size of the ABS is not too large, yet maintaining a high accuracy level. 
So far, we have optimized the ABFs following the above-described procedure for elements of atomic numbers from 2 to 18. 
Below, these newly optimized ABSs are
denoted as ``ABS-opt''. In Table~\ref{tab:1},  the optimized ABSs are compared with the standard ABSs generated by the automatic on-the-fly 
procedure \cite{ren2012resolution} with the NAO-VCC-4Z basis set, in terms of both their size and the obtained RI-RPA correlation energies.  It can be seen that, compared to the standard ABSs generated using the NAO-VCC-4Z, the number of basis functions in the newly optimized
ABS is less, yet yielding better (variationally lower) RPA correlation energies for all elements. 
This study suggests a promising way to create pre-optimized ABSs, which may become advantageous for general molecular or solid-state calculations.

\begin{table}[!h]
\caption{Comparison of the size of two types of ABSs and the corresponding RPA correlation energies obtained from the Sternheimer scheme.
``ABS(NAO-4Z)'' denotes the standard ABS generated via the automatic procedure \cite{ren2012resolution} 
with NAO-VCC-4Z \cite{IgorZhang/etal:2013} basis set; ``ABS-opt'' denotes the ABS generated using the variational optimization 
procedure described in this section, containing exclusively ``hydrogen-like" functions.
 $N_{aux}$ denotes the number of auxiliary function. }
 \label{tab:1}
\begin{tabular}{|c|c|c|c|c|}
\hline \multirow{2}{*}{ Atom } & \multicolumn{2}{c|}{$N_{aux }$} & \multicolumn{2}{c|}{$E_c^{RPA}$ (Ha)} \\
\cline { 2 - 5 } & \text { ABS(NAO-4Z) } & \text { ABS-opt } & \text { ABS (NAO-4Z) } & \text { ABS-opt } \\
\hline {He} & 143 & 98 & -0.0833 & -0.0830 \\
\hline {Li} & 226 & 143 & -0.1087 & -0.1102 \\
\hline {Be} & 234 & 177 & -0.1786 & -0.1788 \\
\hline {B} & 276 & 211 & -0.2229 & -0.2232 \\
\hline {C} & 266 & 238 & -0.2748 & -0.2753 \\
\hline {N} & 280 & 234 & -0.3372 & -0.3372 \\
\hline {O} & 254 & 234 & -0.4102 & -0.4109 \\
\hline {F} & 284 & 259 & -0.4970 & -0.4983 \\
\hline {Ne} & 280 & 278 & -0.5963 & -0.5968 \\
\hline {Na} & 362 & 298 & -0.6137 & -0.6202 \\
\hline {Mg} & 368 & 276 & -0.6751 & -0.6795 \\
\hline {Al} & 388 & 281 & -0.7166 & -0.7213 \\
\hline {Si} & 400 & 295 & -0.7724 & -0.7744 \\
\hline {P} & 381 & 306 & -0.8341 & -0.8373 \\
\hline {S} & 384 & 306 & -0.9063 & -0.9091 \\
\hline {Cl} & 395 & 322 & -0.9876 & -0.9909 \\
\hline {Ar} & 379 & 313 & -1.0792 & -1.0808 \\
\hline
\end{tabular}
\end{table}
\subsection{RPA correlation energies based on the eigenspectrum of $\chi^0 v$}
\label{sec:RPA_eigenspectrum}
As demonstrated above, empowered by the Sternheimer method, as long as high-quality ABS is available, accurate RPA correlation energy 
can be calculated. However, for heavy elements, we do not have NAO-VCC-$n$Z basis sets and hence no corresponding ABSs can be generated at the moment. 
Of course, one can in principle create optimized ABSs according to the procedure described in Sec.~\ref{sec:ABF-opt}. However, for heavy elements, the optimization can get rather expensive and tedious. How to deal with this situation? In fact, as is shown in Ref.~\cite{nguyen2009efficient},
based on the Steinheimer equation, it is possible to compute accurate RPA correlation energies without invoking an ABS.

From Eq.~\ref{eq:EcRPA}, one may realize that the RPA correlation energy can be readily computed once  
the eigenspectra of the operator $\chi^0(i\omega) v$ are determined for a set of frequency grid points. In terms of a set of ABFs, $\chi^0(i\omega) v$
is represented as a matrix; one can obtain the eigenspectrum of $\chi^0(i\omega) v$ by diagonalizing the matrix. 
Alternatively, utilizing the Sternheimer equation, one can employ the 
power iteration method \cite{booth2006power} to determine the eigenspectrum
of the operator $\chi^0(i\omega) v$ iteratively, in terms of which the RPA 
correlation energy can be obtained, without the need of constructing a high-quality ABS.

The machinery of the power iteration method works as follows: One starts with an initial trial wave function $\left | \varphi  \right \rangle $, which, without losing generality for atoms, also has the form of a radial function times spherical harmonics, 
\begin{equation}
\varphi(\bm{r})=\varphi(r)Y_L^M(\theta,\phi)
\end{equation}
Then, applying the operator $\chi^0(i\omega) v$  to $\left | \varphi  \right \rangle $, one can obtain the first-order density $\Delta n(\bm r)$ by solving Sternheimer equation. Formally, one has 
\begin{equation}
\chi^0 v\left | \varphi  \right \rangle= \Delta n\, ,
\label{eq:eigen_equ}
\end{equation}
which physically corresponds to the first-order change of the density of KS system induced by an ``external'' potential $v|\varphi\rangle$,
which is nothing but the Hartree potential associated with $\varphi(\bm{r})$,
\begin{equation}
\langle\bfr | v\varphi\rangle = \int \frac{\varphi(\bfrp)}{|\bfr - \bfrp|} d\bfrp \, .
\end{equation}
Thus, $\Delta n$ in Eq.~\ref{eq:eigen_equ} can be readily computed by solving the Sternheimer equation with $\langle\bfr | v\varphi\rangle$ taken as the perturbation.
As discussed in the previous section, for a spherical atom, the first-order density has the same angular distribution of the perturbed Hamiltonian, namely,
\begin{equation}
\Delta n(\bm r)=\Delta n(r)Y_L^M(\theta,\phi) \, .
\end{equation}
We then normalize the resultant $\Delta n(r)$ and take it as the ``perturbation'' of the next iteration. Repeating this process, one will attain
a right eigenvector of the operator $\chi^0(i\omega) v$ and the corresponding eigenvalue. Next, taking another trial function and 
orthogonalize it against the eigenvectors that have already been found, one will get the next eigenvalue and eigenvector.  
The eigenspectrum of $\chi^0(i\omega) v$ is negative definite, and bounded below zero. 
By performing such operation repeatedly, one can gradually retrieve the  eigenspectrum  of the operator $\chi^0(i\omega) v$ belonging to a given angular momentum channel $L$, up to a pre-set
threshold for the eigenvalues. 
It is worth noting that the different values of $M$ correspond to degenerate eigenvectors, owing to the spherical symmetry of atoms. So the degeneracy of each eigenvalue belonging to a given angular momentum channel $L$ is 2$L$+1. By varying the angular momentum $L$ and repeating the above iteration process, one can 
obtain the entire eigenspectrum of any angular momentum channel, at any imaginary frequency point.
As mentioned above, once the eigenspectrum of $\chi^0(i\omega) v$ is attained, the RPA correlation energy can be trivially calculated
via Eq.~\ref{eq:EcRPA}. Moreover, this method 
does not depend on any basis set. The convergence behavior of this ``basis-free'' approach can be examined from two aspects: First, the convergence  
of the RPA correlation energy with the increase of the number of eigenvalues for a fixed angular momentum channel $L$. Second, the convergence behavior of RPA correlation energy with respect to the maximum value of the angular momentum $L_{max}$. 
A detailed analysis of the convergence behavior with respect to these two factors 
is given in Appendix~\ref{appendixd}, using Kr atom and Ar atom as examples. 

Because no pre-constructed basis set is needed, we can compute the
RPA correlation energy for any element. Another 
important point is that the eigenspectrum obtained via 
the power iteration method is given systematically from large to small in terms of absolute values, 
and their contributions to the RPA correlation energy are hence getting gradually smaller. 
Therefore, for a given $L$,  one can make the calculation arbitrarily accurate by increasing the number of 
eigenvalues included. Here, we set the precision threshold to be 0.1 meV, i.e., further eigenvalues will be discarded if 
the change of the RPA correlation energy is below 0.1 meV. Regarding the convergence behavior with respect to $L_{max}$, we observe
a rather good $1/L_{max}^3$ of the calculated RPA correlation energy for large $L_{max}$ (cf. Appendix~\ref{appendixd}), 
which is consistent with the early analysis based on
the quantum chemistry correlated methods \cite{Schwartz:1962,Hill:1985}. Such a behavior allows one to readily extrapolate the results to 
the limit of $L_{max} \rightarrow \infty$.
In Table~\ref{tab:2} we present the obtained all-electron RPA@LDA correlation energies for atoms of species from H to Kr (1-36). 
Results for both $L_{max}=14$ and extrapolated $L_{max}\rightarrow \infty$ are reported, in comparison to the reference results obtained using the
“hard-wall cavity” method \cite{jiang2007random} with $L_{max}=14$, available only for closed-shell and half-filled open-shell atoms. 
We note that the ``basis-free'' approach
for calculating the RPA correlation energy described here is very similar to that discussed in Ref.~\cite{nguyen2009efficient}, but
there only results for a few closed-shell atoms are presented.

Table~\ref{tab:2} shows that the Sternheimer scheme together with the power iteration method yields results that are in excellent agreement with
the reference results obtained using the ``hard-wall cavity'' method \cite{jiang2007random}. In particular, the results presented in the third column and the reference
results presented in the fourth column are both obtained with $L_{max}=14$ and hence directly comparable. For closed-shell atoms, the difference
between the results yielded by these two rather different approaches are $\sim$ 50 meV or smaller. For half-filled open-shell atoms, the difference is noticeably larger
(at the level of 0.1 eV), and this is mostly because spin degeneracy is assumed in the Sternheimer-based calculations (due to the restriction
imposed by the ``dftatom'' code \cite{vcertik2013dftatom}), where spin-polarized
calculations were done in case of the ``hard-wall cavity'' method. Test RPA calculations based on the usual SOS scheme as implemented in FHI-aims indicates
that the RPA correlation energies obtained with spin-polarized and spin-degenerate DFA configurations have differences at the level of 0.1 eV for most open-shell atoms. Finally, the difference between the second and third columns in Table~\ref{tab:2} reflects the deviation between the result
obtained with a large but finite $L_{max}$ from the extrapolated $L_{max}\rightarrow \infty$ result. The deviation is only about a couple of meV for light elements
and increases to more than 0.1 eV for the heavier (fourth-row) elements. Such a difference has no relevance for any binding energy calculations
that are of physical interest, but nevertheless the second column of Table~\ref{tab:2} provides numerically highly converged all-electron atomic RPA correlation energies for all chemical
elements from 1 to 36, which can serves as reference values for any future studies along this line.
 
\begin{table}[!h]
\caption{RPA@LDA correlation energies (in eV) for atoms  with atomic numbers 1-36. The first column presents the extrapolated 
($L_{max}\rightarrow \infty$) 
basis-free results, and the second column contains results obtained with $L_{max}=14$. The reference results in the third
column are obtained using the ``hard-wall cavity'' method \cite{jiang2007random} with $L_{max}=14$.}
\begin{tabular}{c c c c c}

\hline
\textbf{Atom} & \textbf{RPA-basis-free ($L_{max}\rightarrow \infty$) } & \textbf{$L_{max}=14$}   & \textbf{Reference}\\ \hline
H  &    -0.568    &   -0.568     &    -0.569 \\
He & -2.287 & -2.286 & -2.288 \\
      Li & -3.045 & -3.044 &-3.078 \\ 
      Be & -4.950  & -4.948 & -4.952 \\ 
      B  &  -6.159 & -6.156 &  / \\
      C  &  -7.612 & -7.607  & / \\
      N  &  -9.348 & -9.341 & -9.241 \\
      O  &  -11.403 & -11.393 & / \\
      F  &  -13.801  & -13.789 & / \\
      Ne &  -16.567  &  -16.552 &  -16.517 \\
      Na &  -17.239  &   -17.222& -17.211 \\
      Mg &  -18.929 & -18.911 &  -18.857 \\
      Al &  -20.115 & -20.095 &   / \\
      Si &  -21.610 & -21.587 & /\\
      P  &  -23.381 & -23.355 & -23.219 \\
      S  &  -25.415 & -25.385 & / \\
      Cl &  -27.709 &  -27.675 & / \\
      Ar &  -30.269 &   -30.231 & -30.281 \\
      K &  -31.757 & -31.716 & /\\
      Ca & -34.191 & -34.147 & /\\
      Sc & -36.479  &  -36.431 & /\\
      Ti & -37.306  &  -37.252 & /\\
      V & -40.914   &  -40.854 & /\\
      Cr & -44.348  &  -44.280 & /\\
      Mn & -46.031  &  -45.958 & /\\
      Fe & -48.966  &  -48.885 & /\\
      Co & -52.091 & -52.002 &  /\\
      Ni & -55.410 & -55.313  &  /\\
      Cu  & -61.451 & -61.343  & /\\
      Zn  & -62.649 & -62.533  &  /\\
      Ga  & -63.003  & -62.882 &   /\\
     Ge   & -63.940 &-63.813  &    /\\
      As  & -65.297 &-65.164  &  /\\
      Se & -66.997 &-66.858  &  /\\ 
      Br &-68.992 &-68.845 & /\\
      Kr & -71.291 & -71.136  & /\\
      
      \hline
      \hline
     \label{tab:2}
\end{tabular}
\end{table}

\section{Conclusions}
\label{sec:conclusion}
The convergence of explicit correlated methods in terms of atomic orbitals to the CBS limit is a challenging problem. 
Practical calculations rely on the availability of high-quality hierarchical basis sets and the extrapolation to the CBS limit
based on empirical rules. In this work, we demonstrate that, by solving the Sternheimer equation on a dense radial grid, one is able to
obtain numerically fully converged RI-RPA correlation energy for atoms. It is shown that the RPA correlation energies obtained with
the biggest available atomic orbitals within the usual SOS scheme are still quite far from the converged results
obtained using the Sternheimer scheme. The BSIE of the RI-RPA scheme is analyzed in terms of both the SPBS and ABS, and it is found
that the major source of BSIE arises from the SPBS whereas the error due to ABS is marginal. Our scheme also offers a recipe 
to variationally construct optimal ABS for RI-RPA calculations, and provides insights about the deficiencies of existing
AO basis sets. The latter is important for designing improved AO basis sets.

The numerical technique we developed for solving the Sternheimer equation for atomic RPA calculations can be extended to diatomic molecules
and to other correlated methods like $GW$. Works along these lines are ongoing. Advances in such numerical techniques will not only provide
unprecedented benchmark values for atom and diatomic systems, but also provide valuable guidance for developing high-quality AO (in particular NAO)
basis sets for correlated calculations for general molecular and extended systems.


\section*{Data availability}
Data that support the findings of this study are available from the
corresponding author upon reasonable request.

\begin{acknowledgement}
We thank Christoph Friedrich and Volker Blum for very helpful discussions. 
This work was funded by the National Key Research and Development Program of China (Grant No. 2022YFA1403800), 
the National Natural Science Foundation of China (Grant Nos. 12188101 and 12134012), and the Max Planck Partner Group project 
on \textit{Advanced Electronic Structure Methods}.
\end{acknowledgement}

\begin{appendix}
\section{Frequency-dependent linear response theory}
\label{appendixA}
The basic frequency-dependent linear response theory has been discussed in the literature (e.g., cf. Ref.~\cite{marques2012fundamentals}).
For the paper to be self-contained, we present below a derivation of the frequency-dependent linear response theory to set up the basis equations
and the notational system.
We start with the single-particle KS Hamiltonian of the system,
\begin{equation}
    H^{(0)}(\bm{r})=-\frac{1}{2}\bigtriangledown^2+V_{eff}(\bm{r})
\end{equation}
where $V_{eff}(\bm{r})$ usually contains three terms,
\begin{equation}
    V_{eff}(\bm{r})=V_{ext}(\bm{r})+V_{h}(\bm{r})+V_{xc}(\bm{r})\, .
\end{equation}
Consider a time-dependent perturbed external potential $V^{(1)}(\bm{r},t)$ that has the following form,
\begin{equation}
    V^{(1)}(\bm{r},t)=V^{(1)}(\bm{r})(e^{i\omega t}+e^{-i\omega t})e^{\eta t}
\end{equation}
where $\omega$ represents the frequency of the monochromatic perturbation, and $\eta=0^+$, indicating that the perturbation is adiabatically
switched on from the remote past $t=-\infty$.
Now the full Hamiltonian is given by
\begin{equation}
    H(\bm{r},t)=H^{(0)}(\bm{r})+V^{(1)}(\bm{r},t)
\end{equation}
whose eigenfunctions can be generally written in the form of perturbation expansion,
\begin{equation}
    \psi_i(\bm{r},t)=(\psi^{(0)}_i(\bm{r})+\psi^{(1)}_i(\bm{r},t)+\psi^{(2)}_i(\bm{r},t)+\dots)e^{i(E^{(0)}_i+E^{(1)}_i(t)+E^{(2)}_i(t)+\dots )t}\, .
\end{equation}
In particular, $\psi^{(0)}_i(\bm{r})$ satisfies the stationary (zeroth-order) Schr\"{o}dinger equation:
\begin{equation}
    H^{(0)}(\bm{r})\psi^{(0)}_i(\bm{r})=E^{(0)}_i\psi^{(0)}_i(\bm{r})\, .
    \label{eq:0-order sch equ}
\end{equation}
In linear response theory, one discards second- and higher-order terms, arriving at
\begin{equation}
    \psi_i(\bm{r},t)\approx(\psi^{(0)}_i(\bm{r})+\psi^{(1)}_i(\bm{r},t))e^{i(E^{(0)}_i+E^{(1)}_i(t))t}
    \label{eq:psi}
\end{equation}
Here, the expression of $E^{(1)}_i(t)$ is akin to the stationary perturbation theory,
\begin{equation}
   E^{(1)}_i(t)=\int_{-\infty}^{t}{\rm d}{t^\prime} \left \langle \psi_i^{(0)}\right |V^{(1)}(t^\prime) \left | \psi_i^{(0)} \right \rangle 
   \label{eq:1-order energy}
\end{equation}
In the linear response regime, the first-order wave function should have the same time dependence as the perturbation,
\begin{equation}
    \psi^{(1)}_i(\bm{r},t)=(\psi^{(1)}_{i,+\omega}(\bm{r})e^{i\omega t}+\psi^{(1)}_{i,-\omega}(\bm{r})e^{-i\omega t})e^{\eta t}\, .
    \label{eq:1-order psi}
\end{equation}
Now we need to find the differential equation that $\psi_{i,\pm \omega}^{(1)}$ satisfies. To this end, we start from the time-dependent Schr\"{o}dinger equation,
\begin{equation}
   i\frac{d\psi_i(\bm{r},t)}{d t}=(H^{(0)}(\bm{r})+V^{(1)}(\bm{r},t))\psi_i(\bm{r},t) \, ,
   \label{eq:sch_equ}
\end{equation}
and by approximating $\psi_i(\bm{r},t)$ by its first-order expansion (Eq.~ \ref{eq:psi}), the left-hand side ($l.h.s.$) and right-hand side
($r.h.s.$) of Eq.~ \ref{eq:sch_equ} can be separately expressed as,

\begin{equation}
  \begin{aligned}
   l.h.s &  =i\frac{d((\psi^{(0)}_i(\bm{r})+\psi^{(1)}_i(\bm{r},t)e^{i(E^{(0)}_i+E^{(1)}_i(t))t})}{dt} \\
   r.h.s & =(H^{(0)}(\bm{r})+V^{(1)}(\bm{r},t))\left[(\psi^{(0)}_i(\bm{r})+\psi^{(1)}_i(\bm{r},t)e^{i(E^{(0)}_i+E^{(1)}_i(t))t}\right]
   \label{eq:right}
   \end{aligned}
\end{equation}

Using Eqs.~\ref{eq:1-order energy} and \ref{eq:1-order psi}, one can get,
\begin{equation}
\begin{aligned}
 &\tilde{E}^{(1)}_i\psi^{(0)}_i(\bm{r})(e^{i\omega t}+e^{-i\omega t}) +E^{(0)}_i(\psi^{(1)}_{i,+\omega}(\bm{r})e^{i\omega t}+\psi^{(1)}_{i,-\omega}(\bm{r})e^{-i\omega t})+ \\ 
 &(-\omega+i\eta)\psi^{(1)}_{i,+\omega}(\bm{r})e^{i\omega t}+(\omega+i\eta)\psi^{(1)}_{i,-\omega}(\bm{r})e^{-i\omega t} \\
  = & H^{(0)}(\bm{r})(\psi^{(1)}_{i,+\omega}(\bm{r})e^{i\omega t}+\psi^{(1)}_{i,-\omega}(\bm{r})e^{-i\omega t})+V^{(1)}(\bm{r})\psi^{(0)}_i(\bm{r})(e^{i\omega t}+e^{-i\omega t}) \, ,
  \label{eq:1-order equ}
\end{aligned}
\end{equation}
where $\tilde{E}^{(1)}_i$ is given by,
\begin{equation}  
   \tilde{E}^{(1)}_i=\left \langle \psi_i^{(0)}\right |V^{(1)} \left | \psi_i^{(0)} \right \rangle \, .
\end{equation}
In the derivation of Eq.~\ref{eq:1-order equ}, the second-order terms are discarded, whereas the zero-order term is eliminated by using  Eq.~\ref{eq:0-order sch equ}.

Equation~\ref{eq:1-order equ} can be reorganized into two equations, utilizing the linear independence of $e^{i\omega t}$ and $e^{-i\omega t}$,
\begin{equation}  
   (H^{(0)}(\bm{r})-E^{(0)}_i+\omega-i\eta)\psi^{(1)}_{i,+\omega}(\bm{r})=(\tilde{E}^{(1)}_i-V^{(1)}(\bm{r}))\psi^{(0)}_i(\bm{r})
   \label{eq:+omega}
\end{equation}
\begin{equation}  
   (H^{(0)}(\bm{r})-E^{(0)}_i-\omega-i\eta)\psi^{(1)}_{i,-\omega}(\bm{r})=(\tilde{E}^{(1)}_i-V^{(1)}(\bm{r}))\psi^{(0)}_i(\bm{r})
    \label{eq:-omega}
\end{equation}
In fact, Eq.~\ref{eq:+omega} and Eq.~\ref{eq:-omega} are essentially the same equation,
\begin{equation}  
   (H^{(0)}(\bm{r})-E^{(0)}_i+\omega-i\eta)\psi^{(1)}_{i,\omega}(\bm{r})=(\tilde{E}^{(1)}_i-V^{(1)}(\bm{r}))\psi^{(0)}_i(\bm{r})
   \label{eq:real-frequence st equ}
\end{equation}
which is referred to as the frequency-dependent Sternheimer equation.

In RPA calculations, it is convenient to work with imaginary frequencies.  To this end,
one can analytically continue the frequency variable from real to imaginary domains, 
and Eq.~\ref{eq:real-frequence st equ} then becomes,
\begin{equation}  
   (H^{(0)}(\bm{r})-E^{(0)}_i+i\omega)\psi^{(1)}_{i,i\omega}(\bm{r})=(\tilde{E}^{(1)}_i-V^{(1)}(\bm{r}))\psi^{(0)}_i(\bm{r})\, .
   \label{eq:imaginary-frequence st equ}
\end{equation}
This is the Eq.~\ref{eq:st} in the main text we would like to derive. 

\section{Derivation of radial Sternheimer equation}
\label{appendixb}
Starting from Eq.~\ref{eq:st}, the original frequency-dependent Sternheimer equation can be reduced to an one-dimensional radial equation.
Here we show how this simplification is achieved.
Using Eqs.~\ref{eq:H^0}-\ref{eq:psi^0}, \ref{eq:auxil_P} and \ref{eq:psi_1_comp}, the $l.h.s.$ and $r.h.s.$ of Eq.~\ref{eq:st} become,
\begin{eqnarray}
l.h.s.&=&(-\frac{1}{2}\nabla^2+V_{eff}(r)-\epsilon_i+i\omega)\left[\sum_{l^\prime m^\prime}u_{i\mu,l^\prime m^\prime}^{(1)}(r,i\omega)Y_{l^\prime}^{m^\prime}(\theta ,\phi)\right] 
\label{eq:radial st left1} \\
r.h.s.&=&[\epsilon_i^{(1)}-V_\mu^{(1)}(r)Y_L^M(\theta,\phi)]u_{i,l}(r)Y_l^m(\theta,\phi)
\label{eq:radial st right}
\end{eqnarray}
In terms of spherical coordinates, the kinetic energy operator $-1/2
\nabla^2$ can be decomposed into $-\frac{1}{2r}\frac{\partial^2}{\partial r^2}r+\frac{\widehat{L} ^2}{2r^2}$. And using
\begin{equation}
\widehat{L}^2Y_{l^\prime}^{m^\prime}(\theta,\phi)=l^\prime(l^\prime+1)Y_{l^\prime}^{m^\prime}(\theta,\phi)\, ,
\end{equation}
Eq.~\ref{eq:radial st left1} changes into
\begin{equation}
l.h.s.=\sum_{l^\prime m^\prime}Y_{l^\prime}^{m^\prime}(\theta ,\phi)\left(-\frac{1}{2r}\frac{\partial^2}{\partial r^2}r+\frac{l^\prime(l^\prime+1)}{2r^2}+V_{eff}(r)-\epsilon_i+i\omega\right)u_{i\mu,l^\prime m^\prime}^{(1)}(r,i\omega) \, .
\label{eq:radial st left2}
\end{equation}
Multiplying both sides of the equation with $Y_{l^{\prime\prime}}^{m^{\prime\prime}}(\theta,\phi)$ and integrating with respect to the angular coordinates ${(\theta,\phi)}$, and further utilizing the orthogonality relationship between real spherical harmonic functions,
\begin{equation}
\int{\rm d}{\Omega}{Y_{l^\prime}^{m^\prime}}(\theta,\phi)Y_{l}^{m}(\theta,\phi)=\delta_{ll^{\prime}}\delta_{mm^{\prime}}\, .
\end{equation}
one gets,
\begin{eqnarray}
 l.h.s. &=&\sum_{l^\prime m^\prime}\delta_{l^{\prime \prime}l^{\prime}}\delta_{m^{\prime \prime}m^{\prime}}\left(-\frac{1}{2r}\frac{\partial^2}{\partial r^2}r+\frac{l^\prime(l^\prime+1)}{2r^2}+V_{eff}(r)-\epsilon_i+i\omega\right)u_{i\mu,l^\prime m^\prime}^{(1)}(r,i\omega)  \nonumber \\
 &=&\left(-\frac{1}{2r}\frac{\partial^2}{\partial r^2}r+\frac{l^{\prime \prime}(l^{\prime \prime}+1)}{2r^2}+V_{eff}(r)-\epsilon_i+i\omega\right)u_{i\mu,l^{\prime \prime} m^{\prime \prime}}^{(1)}(r,i\omega)  \label{eq:radial_lhs} \\
 r.h.s. & =&[\epsilon_i^{(1)}\delta_{l^{\prime \prime}l}\delta_{m^{\prime \prime}m}-G_{l^{\prime \prime} Ll}^{m^{\prime \prime} Mm}V_\mu^{(1)}(r)]u_{i,l}(r)
\end{eqnarray}
$\epsilon^{(1)}_i$ in Eq.~\ref{eq:1 order energy 3 dimension} is given by the Gaunt coefficient multiplied by a radial integral.
\begin{equation}
\begin{aligned}
  \epsilon_i^{(1)}&=\left \langle \psi_i\right |V^{(1)} \left | \psi_i \right \rangle \\
  &=G_{lLl}^{mMm}\int{\rm d} {r}u_{i,l}^*(r)V_\mu^{(1)}(r))u_{i,l}(r)
\end{aligned}
\end{equation}
Due to the existence of $\delta_{l^{\prime \prime}l}\delta_{m^{\prime \prime}m}$, the Gaunt coefficient can be moved outside the bracket, 
ending up with
\begin{equation}
\begin{aligned}
 r.h.s.=G_{l^{\prime \prime} Ll}^{m^{\prime \prime} Mm}[\epsilon_{i-radial}^{(1)}\delta_{l^{\prime \prime}l}\delta_{m^{\prime \prime}m}-V_\mu^{(1)}(r)]u_{i,l}(r)
\end{aligned}
\label{eq:radial_rhs}
\end{equation}
Finally, by equating Eq.~\ref{eq:radial_lhs} and \ref{eq:radial_rhs} and replacing$\left \{ {l^{\prime \prime},m^{\prime \prime}} \right \} $ 
by $\left \{ {l^{\prime},m^{\prime}} \right \} $, one obtains the desired radial Sternheimer equation given by Eq.~\ref{eq:radial st}.

\section{Solving the radial Sternheimer equation by finite-difference method}
\label{appendixc}
The radial Sternheimer equation (Eq.~\ref{eq:radial st}) is solved using finite-difference method, which is
implemented in FHI-aims, whereby the radial wavefunctions are tabulated on a logarithmic grid,
\begin{equation}
\label{log grid}
r(i)=r_{min}*r_{inc}^{(i-1)} \, .
\end{equation}
Such coordinates are more suitable for describing the wave functions near the nucleus, but uneven grids are not convenient for 
the implementation of finite-difference method. So we introduce coordinate transformation to convert logarithmic coordinate $r(i)$ to uniform coordinate $x(i)$, and the corresponding relationship is as follows,
\begin{equation}
\begin{aligned}
\label{uniform grid}
&x(i)=ln(r(i))=ln(r_{min})+(i-1)*ln(r_{inc}) \\
&\frac{d}{dr}=\frac{dx}{dr}\frac{d}{dx}=\frac{1}{r}\frac{d}{dx}=e^{-x}\frac{d}{dx}
\end{aligned}
\end{equation}

The key issue is the treatment of second derivative in Eqs.~\ref{eq:radial st} and \ref{eq:radial st degenrate}. Introduce $f(r)=\sqrt{r}u^{(1)}(r)$, we can get,
\begin{equation}
\begin{aligned}
\frac{1}{r}\frac{\partial^2}{\partial r^2}\,ru^{(1)}(r)&=\frac{1}{r}\frac{\partial}{\partial r}\left(\frac{\partial}{\partial r}\sqrt{r}f(r)\right)\\
&=\frac{1}{r^2}\frac{\partial}{\partial x}e^{-x}\frac{\partial}{\partial x}e^{\frac{1}{2}x}f(x)\\
&=r^{-5/2} (f^{\prime\prime}(x)-\frac{1}{4}f(x))
\end{aligned}
\end{equation}
On uniform grids, $f^{\prime \prime}(x)$ can be expressed as:
\begin{equation}
f^{\prime \prime}[x(i)]=\frac{f[x(i+1)]+f[x(i-1)]-2f[x(i)]}{\Delta x^2}
\end{equation}
Therefore, the expression of the second-order differential equation (taking Eq.~\ref{eq:radial st} as an example, and Eq.~\ref{eq:radial st degenrate} is the same) on the grids is given as follows,
\begin{equation}
\begin{aligned}
\label{eq:finite difference radial equation}
&-\frac{1}{2}\frac{f[x(i+1)]+f[x(i-1)]-2f[x(i)]}{\Delta x^2}+\frac{(l^\prime+1/2)^2}{2}+r[x(i)]^2(V_{eff}[x(i)]-\epsilon_i+i\omega)f[x(i)]\\
=\,&G_{l^\prime Ll}^{m^\prime Mm}r[x(i)]^{5/2}\left(\epsilon_{i-rad}^{(1)}\delta_{ll^\prime}\delta_{mm^\prime}-V_\mu^{(1)}[x(i)]\right)u_{i,l}[x(i)]
\end{aligned}
\end{equation}
Equation~\ref{eq:finite difference radial equation} can be transformed into a linear system of equations $Af=B$.
$A$ has the following tri-diagonal symmetric form:
\begin{equation}
A=\left(\begin{array}{cccc}
				\frac{1}{\Delta x^2}+\frac{(l^{\prime}+0.5)^2}{2}+r(1)^2(v_{eff}(1)-\epsilon_i+i\omega) & -\frac{1}{2\Delta x^2}                                                                  & 0                & ... \\
				-\frac{1}{2\Delta x^2}                                                                  & \frac{1}{\Delta x^2}+\frac{(l^{\prime}+0.5)^2}{2}+r(2)^2(v_{eff}(2)-\epsilon_i+i\omega) & -\frac{1}{2\Delta x^2} & ... \\
				0                                                                                 & -\frac{1}{2\Delta x^2}                                                                  & ...              & ...\\
                 ... &...&...&...
			\end{array}\right)
\end{equation}

\section{Convergence behavior analysis}
\label{appendixd}
\subsection{Convergence behavior with respect to the eigenvalues of $\chi^0 v$}
\label{appendixda}
In Sec.~\ref{sec:RPA_eigenspectrum}, we discussed how to calculate the RPA correlation energy in terms of the eigenspectrum of the $\chi^0 v$
operator. Under a certain angular momentum truncation $L_{max}$, we can gradually increase the number of eigenvalues for each angular momentum channel, and monitor the convergence behavior of the RPA correlation energy. It should be noted that the eigenspectrum of  $\chi^0 v$ determined via the power iteration method is given in descending order in term of the absolute value of the eigenvalues, and hence the change in RPA correlation energy will become gradually smaller. We find that, if the change of the RPA correlation energy  $\Delta E_c^{RPA}$ is less than 0.1 meV upon including one more eigenvalue, the RPA correlation energy can converge to within 1 meV. Taking the Kr atom as an example, we found that, starting from the 30th eigenvalue on, each eigenvalue contributes less than 0.1 meV to the RPA correlation energy. Figure 1 shows the cumulative contributions stemming from the $30th$
to $100th$ eigenvalues to the RPA correlation energy. As can be clearly seen, the cumulative energy error is below 1 meV by increasing the number of eigenvalues from 30 to 100.  This means that the remaining error resulting from
the convergence threshold we set (below 0.1 meV upon including one more eigenvalue) is minute.
 \begin{figure}[htbp]
    \centering
    \subfloat{\includegraphics[scale=0.5]{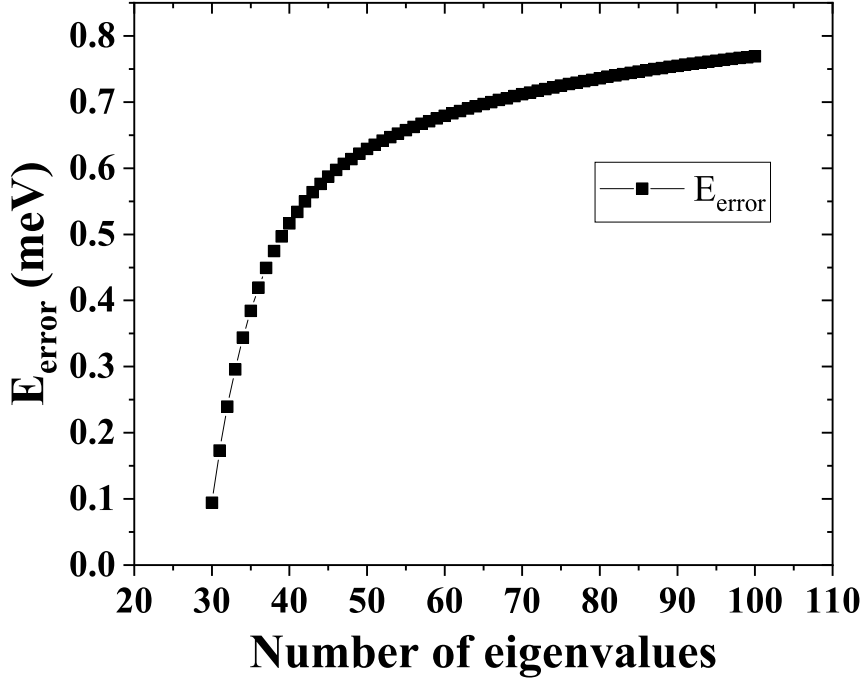}}
    \caption{The change of the RPA correlation energy of the Kr atom by increasing the numbers of eigenvalues of $\chi^0v$ in the calculation. 
    The figure shows that the cumulative change of the energy is below 1 meV when the number of eigenvalues increases from 30 to 100.
    }
    \label{fig:converge_eigenvalue}
\end{figure}
\\

\subsection{Convergence behavior with respect to $L_{max}$} 
\begin{figure}[htbp]
    \centering
    \subfloat{\includegraphics[scale=0.5]{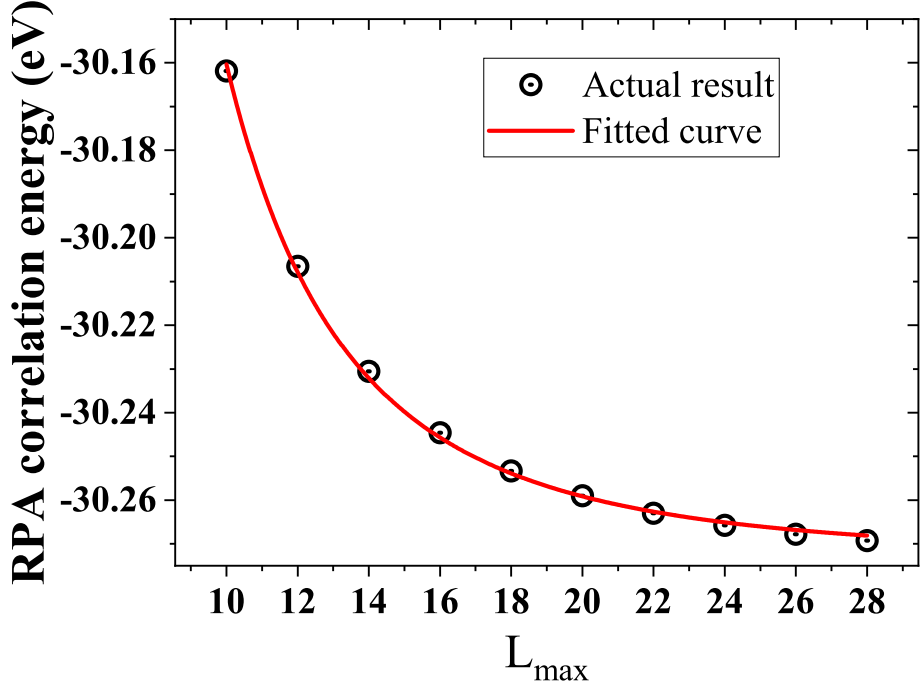}}
    \caption{The convergence of the RPA@LDA correlation energy with respect to the highest angular momentum $L_{max}$ included in the calculation.
    }
    \label{fig:fitted}
 \end{figure}

Early analysis showed that the electron correlation energy for a spherical system converges as $1/L_{max}^3$ where $L_{max}$ is the highest angular momentum included in the calculation \cite{Schwartz:1962,Hill:1985}. Here, we show that the convergence of the RPA correlation energy with respect to $L_{max}$ also follows $1/L_{max}^3$ behavior for large $L_{max}$. An illustrating example is shown in Fig.~\ref{fig:fitted} for the example of Ar atom.
Based on this, we can establish the relationship between the RPA correlation energy under a certain $L_{max}$ truncation and that when $L_{max}$  approaches infinity,
   \begin{equation}
    E_c^{RPA}(L_{max})=E_c^{RPA}(L_{max}\rightarrow {\infty})+\frac{C}{L_{max}^3}
    \label{eq:Ec_lmax}
   \end{equation}
This means that we can fit data obtained with finite $L_{max}$, and obtain $E_c^{RPA}(L_{max}\rightarrow {\infty})$ from an extrapolation.

In Fig.~\ref{fig:fitted}, we presented the RPA correlation energy of Ar atom for $L_{max}$ from 10 to 28. The convergence with respect
to the number of eigenvalues of $\chi^0 v$ under each angular momentum follows the procedure described in Sec.~\ref{appendixda}. We fitted the data according to the expression given in Eq.~\ref{eq:Ec_lmax}, and the fitted curve is shown in Fig.~\ref{fig:fitted}.
The fitting coefficient of determination $R^2$ equals 0.99898, indicating that the fitted data is highly consistent with the fitting expression. The analytical expression obtained from the fitting is given by,
\begin{equation}
    E_c^{RPA}(L_{max})=-30.27325+\frac{112.93756}{L_{max}^3}
\end{equation}
Therefore, the estimated value of $E_c^{RPA}(L_{max}\rightarrow {\infty})$ is $-30.27325$ eV. If we just use three data at $L_{max}=10,12,14$, then the fitted result shows the estimated value of $E_c^{RPA}(L_{max}\rightarrow {\infty})$ is -30.26949 eV. So the error bars of fitting the results with these three data points are within 10 meV.
The presented $L_{max}\rightarrow \infty$ results in Table~\ref{tab:2} are obtained by fitting the data points with $L_{max}=10,12,14$.
\end{appendix}

\bibliography{RPA_bib,CommonBib}

\end{document}

%% file: newcommands.tex
\newcommand{\kpe}{\mathbf{k}\!\cdot\!\mathbf{p}\,}
\newcommand{\Kpe}{\mathbf{K}\!\cdot\!\mathbf{p}\,}
\newcommand{\bfr}{ {\mathbf r}} 
\newcommand{\bfrp}{ {\mathbf r'}} 
\newcommand{\tp}{ {t^\prime}} 
\newcommand{\bfrpp}{ {\bf r^{\prime\prime}}} 
\newcommand{\bfR}{ {\bf R}} 
\newcommand{\bfq}{ {\bf q}} 
\newcommand{\bfp}{ {\bf p}} 
\newcommand{\bfk}{ {\bf k}} 
\newcommand{\bfG}{ {\bf G}} 
\newcommand{\bfGp}{ {\bf G'}} 
\newcommand{\dv}{ {\Delta \hat{v}}} 
\newcommand{\sigmap}{\sigma^\prime} 
\newcommand{\omegap}{\omega^\prime} 
\newcommand{\omegapp}{\omega^{\prime\prime}} 
\newcommand{\bracketm}[1]{\ensuremath{\langle #1   \rangle}}
\newcommand{\bracketw}[2]{\ensuremath{\langle #1 | #2  \rangle}}
\newcommand{\bracket}[3]{\ensuremath{\langle #1 | #2 | #3 \rangle}}
\newcommand{\ket}[1]{\ensuremath{| #1 \rangle}}
\newcommand{\GnWn}{\ensuremath{G_0W_0}\,}
\newcommand{\tn}[1]{\textnormal{#1}}
\newcommand{\f}[1]{\footnotemark[#1]}
\newcommand{\mc}[2]{\multicolumn{1}{#1}{#2}}
\newcommand{\mcs}[3]{\multicolumn{#1}{#2}{#3}}
\newcommand{\mcc}[1]{\multicolumn{1}{c}{#1}}
\newcommand{\refeq}[1]{(\ref{#1})} 
\newcommand{\refcite}[1]{Ref.~\cite{#1}} 
\newcommand{\refsec}[1]{Sec.~\ref{#1}} 
\newcommand\opd{d}
\newcommand\im{i}
\def\bra#1{\mathinner{\langle{#1}|}}
\def\ket#1{\mathinner{|{#1}\rangle}}
\newcommand{\braket}[2]{\langle #1|#2\rangle}
\def\Bra#1{\left<#1\right|}
\def\Ket#1{\left|#1\right>}

\newcommand{\fatr}{\mathbf{r}}
\newcommand{\bmphi}{\bm{\phi}}

%% file: RPA_Sternheimer.bbl
\providecommand{\latin}[1]{#1}
\makeatletter
\providecommand{\doi}
  {\begingroup\let\do\@makeother\dospecials
  \catcode`\{=1 \catcode`\}=2 \doi@aux}
\providecommand{\doi@aux}[1]{\endgroup\texttt{#1}}
\makeatother
\providecommand*\mcitethebibliography{\thebibliography}
\csname @ifundefined\endcsname{endmcitethebibliography}
  {\let\endmcitethebibliography\endthebibliography}{}
\begin{mcitethebibliography}{46}
\providecommand*\natexlab[1]{#1}
\providecommand*\mciteSetBstSublistMode[1]{}
\providecommand*\mciteSetBstMaxWidthForm[2]{}
\providecommand*\mciteBstWouldAddEndPuncttrue
  {\def\EndOfBibitem{\unskip.}}
\providecommand*\mciteBstWouldAddEndPunctfalse
  {\let\EndOfBibitem\relax}
\providecommand*\mciteSetBstMidEndSepPunct[3]{}
\providecommand*\mciteSetBstSublistLabelBeginEnd[3]{}
\providecommand*\EndOfBibitem{}
\mciteSetBstSublistMode{f}
\mciteSetBstMaxWidthForm{subitem}{(\alph{mcitesubitemcount})}
\mciteSetBstSublistLabelBeginEnd
  {\mcitemaxwidthsubitemform\space}
  {\relax}
  {\relax}

\bibitem[T.~H.~Dunning(1989)]{Dunning:1989}
T.~H.~Dunning,~J. Gaussian basis sets for use in correlated molecular
  calculations. I. The atoms boron through neon and hydrogen. \emph{J. Chem.
  Phys.} \textbf{1989}, \emph{90}, 1007\relax
\mciteBstWouldAddEndPuncttrue
\mciteSetBstMidEndSepPunct{\mcitedefaultmidpunct}
{\mcitedefaultendpunct}{\mcitedefaultseppunct}\relax
\EndOfBibitem
\bibitem[Zhang \latin{et~al.}(2013)Zhang, Ren, Rinke, Blum, and
  Scheffler]{IgorZhang/etal:2013}
Zhang,~I.~Y.; Ren,~X.; Rinke,~P.; Blum,~V.; Scheffler,~M. Numeric
  atom-centered-orbital basis sets with valence-correlation consistency from H
  to Ar. \emph{New J. of Phys.} \textbf{2013}, \emph{15}, 123033\relax
\mciteBstWouldAddEndPuncttrue
\mciteSetBstMidEndSepPunct{\mcitedefaultmidpunct}
{\mcitedefaultendpunct}{\mcitedefaultseppunct}\relax
\EndOfBibitem
\bibitem[Helgaker \latin{et~al.}(1997)Helgaker, Koch, and
  Noga]{Helgaker/etal:1997}
Helgaker,~T.; Koch,~W. K.~H.; Noga,~J. Basis-set convergence of correlated
  calculations on water. \emph{J. Chem. Phys.} \textbf{1997}, \emph{106},
  9639\relax
\mciteBstWouldAddEndPuncttrue
\mciteSetBstMidEndSepPunct{\mcitedefaultmidpunct}
{\mcitedefaultendpunct}{\mcitedefaultseppunct}\relax
\EndOfBibitem
\bibitem[Eshuis \latin{et~al.}(2012)Eshuis, Bates, and
  Furche]{Eshuis/Bates/Furche:2012}
Eshuis,~H.; Bates,~J.~E.; Furche,~F. Electron Correlation Methods Based on the
  Random Phase Approximation. \emph{Theor. Chem. Acc.} \textbf{2012},
  \emph{131}, 1084\relax
\mciteBstWouldAddEndPuncttrue
\mciteSetBstMidEndSepPunct{\mcitedefaultmidpunct}
{\mcitedefaultendpunct}{\mcitedefaultseppunct}\relax
\EndOfBibitem
\bibitem[Ren \latin{et~al.}(2012)Ren, Rinke, Joas, and
  Scheffler]{Ren/etal:2012b}
Ren,~X.; Rinke,~P.; Joas,~C.; Scheffler,~M. Random-phase approximation and its
  applications in computational chemistry and materials science. \emph{J.
  Mater. Sci.} \textbf{2012}, \emph{47}, 7447\relax
\mciteBstWouldAddEndPuncttrue
\mciteSetBstMidEndSepPunct{\mcitedefaultmidpunct}
{\mcitedefaultendpunct}{\mcitedefaultseppunct}\relax
\EndOfBibitem
\bibitem[Zhang and Ren(2023)Zhang, and Ren]{Zhang/Ren:2023}
Zhang,~I.~Y.; Ren,~X. Introduction to the Fifth-rung Density Functional
  Approximations: Concept, Formulation, and Applications. \textbf{2023},
  arXiv:2301.12119\relax
\mciteBstWouldAddEndPuncttrue
\mciteSetBstMidEndSepPunct{\mcitedefaultmidpunct}
{\mcitedefaultendpunct}{\mcitedefaultseppunct}\relax
\EndOfBibitem
\bibitem[Jensen \latin{et~al.}(2017)Jensen, Saha, Flores-Livas, Huhn, Blum,
  Goedecker, and Frediani]{Jensen/etal:2017}
Jensen,~S.~R.; Saha,~S.; Flores-Livas,~J.~A.; Huhn,~W.; Blum,~V.;
  Goedecker,~S.; Frediani,~L. The Elephant in the Room of Density Functional
  Theory Calculations. \emph{The Journal of Physical Chemistry Letters}
  \textbf{2017}, \emph{8}, 1449--1457, PMID: 28291362\relax
\mciteBstWouldAddEndPuncttrue
\mciteSetBstMidEndSepPunct{\mcitedefaultmidpunct}
{\mcitedefaultendpunct}{\mcitedefaultseppunct}\relax
\EndOfBibitem
\bibitem[Sternheimer(1951)]{Sternheimer:1951}
Sternheimer,~R.~M. On Nuclear Quadrupole Moments. \emph{Phys. Rev.}
  \textbf{1951}, \emph{84}, 244\relax
\mciteBstWouldAddEndPuncttrue
\mciteSetBstMidEndSepPunct{\mcitedefaultmidpunct}
{\mcitedefaultendpunct}{\mcitedefaultseppunct}\relax
\EndOfBibitem
\bibitem[Sternheimer(1954)]{sternheimer1954electronic}
Sternheimer,~R.~M. Electronic polarizabilities of ions from the Hartree-Fock
  wave functions. \emph{Physical Review} \textbf{1954}, \emph{96}, 951\relax
\mciteBstWouldAddEndPuncttrue
\mciteSetBstMidEndSepPunct{\mcitedefaultmidpunct}
{\mcitedefaultendpunct}{\mcitedefaultseppunct}\relax
\EndOfBibitem
\bibitem[Sternheimer(1957)]{sternheimer1957electronic}
Sternheimer,~R.~M. Electronic polarizabilities of ions. \emph{Physical Review}
  \textbf{1957}, \emph{107}, 1565\relax
\mciteBstWouldAddEndPuncttrue
\mciteSetBstMidEndSepPunct{\mcitedefaultmidpunct}
{\mcitedefaultendpunct}{\mcitedefaultseppunct}\relax
\EndOfBibitem
\bibitem[Sternheimer(1970)]{sternheimer1970quadrupole}
Sternheimer,~R.~M. Quadrupole polarizabilities of various ions and the alkali
  atoms. \emph{Physical Review A} \textbf{1970}, \emph{1}, 321\relax
\mciteBstWouldAddEndPuncttrue
\mciteSetBstMidEndSepPunct{\mcitedefaultmidpunct}
{\mcitedefaultendpunct}{\mcitedefaultseppunct}\relax
\EndOfBibitem
\bibitem[Mahan(1980)]{mahan1980modified}
Mahan,~G. Modified Sternheimer equation for polarizability. \emph{Physical
  Review A} \textbf{1980}, \emph{22}, 1780\relax
\mciteBstWouldAddEndPuncttrue
\mciteSetBstMidEndSepPunct{\mcitedefaultmidpunct}
{\mcitedefaultendpunct}{\mcitedefaultseppunct}\relax
\EndOfBibitem
\bibitem[Zangwill and Soven(1980)Zangwill, and Soven]{zangwill1980density}
Zangwill,~A.; Soven,~P. Density-functional approach to local-field effects in
  finite systems: Photoabsorption in the rare gases. \emph{Physical Review A}
  \textbf{1980}, \emph{21}, 1561\relax
\mciteBstWouldAddEndPuncttrue
\mciteSetBstMidEndSepPunct{\mcitedefaultmidpunct}
{\mcitedefaultendpunct}{\mcitedefaultseppunct}\relax
\EndOfBibitem
\bibitem[Mahan(1982)]{mahan1982van}
Mahan,~G. Van der Waals coefficient between closed shell ions. \emph{The
  Journal of Chemical Physics} \textbf{1982}, \emph{76}, 493--497\relax
\mciteBstWouldAddEndPuncttrue
\mciteSetBstMidEndSepPunct{\mcitedefaultmidpunct}
{\mcitedefaultendpunct}{\mcitedefaultseppunct}\relax
\EndOfBibitem
\bibitem[Andrade \latin{et~al.}(2007)Andrade, Botti, Marques, and
  Rubio]{Andrade/etal:2007}
Andrade,~X.; Botti,~S.; Marques,~M. A.~L.; Rubio,~A. Time-dependent density
  functional theory scheme for efficient calculations of dynamic
  (hyper)polarizabilities. \emph{The Journal of Chemical Physics}
  \textbf{2007}, \emph{126}, 184106\relax
\mciteBstWouldAddEndPuncttrue
\mciteSetBstMidEndSepPunct{\mcitedefaultmidpunct}
{\mcitedefaultendpunct}{\mcitedefaultseppunct}\relax
\EndOfBibitem
\bibitem[Gerratt and Mills(1968)Gerratt, and Mills]{Gerratt/Mills:1968}
Gerratt,~J.; Mills,~I.~M. Force Constants and Dipole‐Moment Derivatives of
  Molecules from Perturbed Hartree–Fock Calculations. I. \emph{The Journal of
  Chemical Physics} \textbf{1968}, \emph{49}, 1719--1729\relax
\mciteBstWouldAddEndPuncttrue
\mciteSetBstMidEndSepPunct{\mcitedefaultmidpunct}
{\mcitedefaultendpunct}{\mcitedefaultseppunct}\relax
\EndOfBibitem
\bibitem[Baroni \latin{et~al.}(1987)Baroni, Giannozzi, and
  Testa]{Baroni/etal:1987}
Baroni,~S.; Giannozzi,~P.; Testa,~A. Green's-function approach to linear
  response in solids. \emph{Phys. Rev. Lett.} \textbf{1987}, \emph{58},
  1861--1864\relax
\mciteBstWouldAddEndPuncttrue
\mciteSetBstMidEndSepPunct{\mcitedefaultmidpunct}
{\mcitedefaultendpunct}{\mcitedefaultseppunct}\relax
\EndOfBibitem
\bibitem[Baroni \latin{et~al.}(2001)Baroni, {de Gironcoli}, Corso, and
  Giannozzi]{Baroni/etal:2001}
Baroni,~S.; {de Gironcoli},~S.; Corso,~A.~D.; Giannozzi,~P. Phonons and related
  crystal properties from density-functional perturbation theory. \emph{Rev.
  Mol. Phys.} \textbf{2001}, \emph{73}, 515\relax
\mciteBstWouldAddEndPuncttrue
\mciteSetBstMidEndSepPunct{\mcitedefaultmidpunct}
{\mcitedefaultendpunct}{\mcitedefaultseppunct}\relax
\EndOfBibitem
\bibitem[Gonze(1995)]{Gonze:1995}
Gonze,~X. Adiabatic density-functional perturbation theory. \emph{Phys. Rev. A}
  \textbf{1995}, \emph{52}, 1096--1114\relax
\mciteBstWouldAddEndPuncttrue
\mciteSetBstMidEndSepPunct{\mcitedefaultmidpunct}
{\mcitedefaultendpunct}{\mcitedefaultseppunct}\relax
\EndOfBibitem
\bibitem[Wilson \latin{et~al.}(2008)Wilson, Gygi, and
  Galli]{wilson2008efficient}
Wilson,~H.~F.; Gygi,~F.; Galli,~G. Efficient iterative method for calculations
  of dielectric matrices. \emph{Physical Review B} \textbf{2008}, \emph{78},
  113303\relax
\mciteBstWouldAddEndPuncttrue
\mciteSetBstMidEndSepPunct{\mcitedefaultmidpunct}
{\mcitedefaultendpunct}{\mcitedefaultseppunct}\relax
\EndOfBibitem
\bibitem[Wilson \latin{et~al.}(2009)Wilson, Lu, Gygi, and
  Galli]{wilson2009iterative}
Wilson,~H.~F.; Lu,~D.; Gygi,~F.; Galli,~G. Iterative calculations of dielectric
  eigenvalue spectra. \emph{Physical Review B} \textbf{2009}, \emph{79},
  245106\relax
\mciteBstWouldAddEndPuncttrue
\mciteSetBstMidEndSepPunct{\mcitedefaultmidpunct}
{\mcitedefaultendpunct}{\mcitedefaultseppunct}\relax
\EndOfBibitem
\bibitem[Nguyen and de~Gironcoli(2009)Nguyen, and
  de~Gironcoli]{nguyen2009efficient}
Nguyen,~H.-V.; de~Gironcoli,~S. Efficient calculation of exact exchange and RPA
  correlation energies in the adiabatic-connection fluctuation-dissipation
  theory. \emph{Physical Review B} \textbf{2009}, \emph{79}, 205114\relax
\mciteBstWouldAddEndPuncttrue
\mciteSetBstMidEndSepPunct{\mcitedefaultmidpunct}
{\mcitedefaultendpunct}{\mcitedefaultseppunct}\relax
\EndOfBibitem
\bibitem[Umari \latin{et~al.}(2009)Umari, Stenuit, and
  Baroni]{Umari/Stenuit/Baroni:2009}
Umari,~P.; Stenuit,~G.; Baroni,~S. Optimal representation of the polarization
  propagator for large-scale GW calculations. \emph{Phys. Rev. B}
  \textbf{2009}, \emph{79}, 201104\relax
\mciteBstWouldAddEndPuncttrue
\mciteSetBstMidEndSepPunct{\mcitedefaultmidpunct}
{\mcitedefaultendpunct}{\mcitedefaultseppunct}\relax
\EndOfBibitem
\bibitem[Umari \latin{et~al.}(2010)Umari, Stenuit, and
  Baroni]{Umari/Stenuit/Baroni:2010}
Umari,~P.; Stenuit,~G.; Baroni,~S. GW quasiparticle spectra from occupied
  states only. \emph{Phys. Rev. B} \textbf{2010}, \emph{81}, 115104\relax
\mciteBstWouldAddEndPuncttrue
\mciteSetBstMidEndSepPunct{\mcitedefaultmidpunct}
{\mcitedefaultendpunct}{\mcitedefaultseppunct}\relax
\EndOfBibitem
\bibitem[Giustino \latin{et~al.}(2010)Giustino, Cohen, and
  Louie]{Giustino/Cohen/Louie:2010}
Giustino,~F.; Cohen,~M.~L.; Louie,~S.~G. GW method with the self-consistent
  Sternheimer equation. \emph{Phys. Rev. B} \textbf{2010}, \emph{81},
  115105\relax
\mciteBstWouldAddEndPuncttrue
\mciteSetBstMidEndSepPunct{\mcitedefaultmidpunct}
{\mcitedefaultendpunct}{\mcitedefaultseppunct}\relax
\EndOfBibitem
\bibitem[Nguyen \latin{et~al.}(2012)Nguyen, Pham, Rocca, and
  Galli]{Nguyen/etal:2012}
Nguyen,~H.-V.; Pham,~T.~A.; Rocca,~D.; Galli,~G. Improving accuracy and
  efficiency of calculations of photoemission spectra within the many-body
  perturbation theory. \emph{Phys. Rev. B} \textbf{2012}, \emph{85},
  081101\relax
\mciteBstWouldAddEndPuncttrue
\mciteSetBstMidEndSepPunct{\mcitedefaultmidpunct}
{\mcitedefaultendpunct}{\mcitedefaultseppunct}\relax
\EndOfBibitem
\bibitem[Betzinger \latin{et~al.}(2012)Betzinger, Friedrich, G{\"o}rling, and
  Bl{\"u}gel]{betzinger2012precise}
Betzinger,~M.; Friedrich,~C.; G{\"o}rling,~A.; Bl{\"u}gel,~S. Precise response
  functions in all-electron methods: Application to the
  optimized-effective-potential approach. \emph{Physical Review B}
  \textbf{2012}, \emph{85}, 245124\relax
\mciteBstWouldAddEndPuncttrue
\mciteSetBstMidEndSepPunct{\mcitedefaultmidpunct}
{\mcitedefaultendpunct}{\mcitedefaultseppunct}\relax
\EndOfBibitem
\bibitem[Betzinger \latin{et~al.}(2013)Betzinger, Friedrich, and
  Bl{\"u}gel]{betzinger2013precise}
Betzinger,~M.; Friedrich,~C.; Bl{\"u}gel,~S. Precise response functions in
  all-electron methods: Generalization to nonspherical perturbations and
  application to NiO. \emph{Physical Review B} \textbf{2013}, \emph{88},
  075130\relax
\mciteBstWouldAddEndPuncttrue
\mciteSetBstMidEndSepPunct{\mcitedefaultmidpunct}
{\mcitedefaultendpunct}{\mcitedefaultseppunct}\relax
\EndOfBibitem
\bibitem[Betzinger \latin{et~al.}(2015)Betzinger, Friedrich, G{\"o}rling, and
  Bl{\"u}gel]{betzinger2015precise}
Betzinger,~M.; Friedrich,~C.; G{\"o}rling,~A.; Bl{\"u}gel,~S. Precise
  all-electron dynamical response functions: Application to COHSEX and the RPA
  correlation energy. \emph{Physical Review B} \textbf{2015}, \emph{92},
  245101\relax
\mciteBstWouldAddEndPuncttrue
\mciteSetBstMidEndSepPunct{\mcitedefaultmidpunct}
{\mcitedefaultendpunct}{\mcitedefaultseppunct}\relax
\EndOfBibitem
\bibitem[Gunnarsson and Lundqvist(1976)Gunnarsson, and
  Lundqvist]{Gunnarsson/Lundqvist:1976}
Gunnarsson,~O.; Lundqvist,~B.~I. Exchange and correlation in atoms, molecules,
  and solids by the spin-density-functional formalism. \emph{Phys.\ Rev.\ B}
  \textbf{1976}, \emph{13}, 4274\relax
\mciteBstWouldAddEndPuncttrue
\mciteSetBstMidEndSepPunct{\mcitedefaultmidpunct}
{\mcitedefaultendpunct}{\mcitedefaultseppunct}\relax
\EndOfBibitem
\bibitem[Langreth and Perdew(1977)Langreth, and Perdew]{Langreth/Perdew:1977}
Langreth,~D.~C.; Perdew,~J.~P. Exchange-correlation energy of a metal surface:
  Wave-vector analysis. \emph{Phys. Rev. B} \textbf{1977}, \emph{15},
  2884\relax
\mciteBstWouldAddEndPuncttrue
\mciteSetBstMidEndSepPunct{\mcitedefaultmidpunct}
{\mcitedefaultendpunct}{\mcitedefaultseppunct}\relax
\EndOfBibitem
\bibitem[Marques \latin{et~al.}(2012)Marques, Maitra, Nogueira, Gross, and
  Rubio]{marques2012fundamentals}
Marques,~M.~A.; Maitra,~N.~T.; Nogueira,~F.~M.; Gross,~E.~K.; Rubio,~A.
  \emph{Fundamentals of time-dependent density functional theory}; Springer,
  2012; Vol. 837\relax
\mciteBstWouldAddEndPuncttrue
\mciteSetBstMidEndSepPunct{\mcitedefaultmidpunct}
{\mcitedefaultendpunct}{\mcitedefaultseppunct}\relax
\EndOfBibitem
\bibitem[Talman(2011)]{talman2011multipole}
Talman,~J.~D. Multipole expansions of orbital products about an intermediate
  center. \emph{International Journal of Quantum Chemistry} \textbf{2011},
  \emph{111}, 2221--2227\relax
\mciteBstWouldAddEndPuncttrue
\mciteSetBstMidEndSepPunct{\mcitedefaultmidpunct}
{\mcitedefaultendpunct}{\mcitedefaultseppunct}\relax
\EndOfBibitem
\bibitem[Ren \latin{et~al.}(2012)Ren, Rinke, Blum, Wieferink, Tkatchenko,
  Sanfilippo, Reuter, and Scheffler]{ren2012resolution}
Ren,~X.; Rinke,~P.; Blum,~V.; Wieferink,~J.; Tkatchenko,~A.; Sanfilippo,~A.;
  Reuter,~K.; Scheffler,~M. Resolution-of-identity approach to Hartree--Fock,
  hybrid density functionals, RPA, MP2 and GW with numeric atom-centered
  orbital basis functions. \emph{New Journal of Physics} \textbf{2012},
  \emph{14}, 053020\relax
\mciteBstWouldAddEndPuncttrue
\mciteSetBstMidEndSepPunct{\mcitedefaultmidpunct}
{\mcitedefaultendpunct}{\mcitedefaultseppunct}\relax
\EndOfBibitem
\bibitem[{\v{C}}ert{\'\i}k \latin{et~al.}(2013){\v{C}}ert{\'\i}k, Pask, and
  Vack{\'a}{\v{r}}]{vcertik2013dftatom}
{\v{C}}ert{\'\i}k,~O.; Pask,~J.~E.; Vack{\'a}{\v{r}},~J. dftatom: A robust and
  general Schr{\"o}dinger and Dirac solver for atomic structure calculations.
  \emph{Computer Physics Communications} \textbf{2013}, \emph{184},
  1777--1791\relax
\mciteBstWouldAddEndPuncttrue
\mciteSetBstMidEndSepPunct{\mcitedefaultmidpunct}
{\mcitedefaultendpunct}{\mcitedefaultseppunct}\relax
\EndOfBibitem
\bibitem[Press \latin{et~al.}(2007)Press, Teukolsky, Vetterling, and
  Flannery]{press2007numerical}
Press,~W.~H.; Teukolsky,~S.~A.; Vetterling,~W.~T.; Flannery,~B.~P.
  \emph{Numerical recipes 3rd edition: The art of scientific computing};
  Cambridge university press, 2007\relax
\mciteBstWouldAddEndPuncttrue
\mciteSetBstMidEndSepPunct{\mcitedefaultmidpunct}
{\mcitedefaultendpunct}{\mcitedefaultseppunct}\relax
\EndOfBibitem
\bibitem[Klime\v{s} \latin{et~al.}(2014)Klime\v{s}, Kaltak, and
  Kresse]{Klimes/Kaltak/Kresse:2014}
Klime\v{s},~J.~K.; Kaltak,~M.; Kresse,~G. Predictive GW calculations using
  plane waves and pseudopotentials. \emph{Phys. Phys. B} \textbf{2014},
  \emph{90}, 075125\relax
\mciteBstWouldAddEndPuncttrue
\mciteSetBstMidEndSepPunct{\mcitedefaultmidpunct}
{\mcitedefaultendpunct}{\mcitedefaultseppunct}\relax
\EndOfBibitem
\bibitem[Ihrig \latin{et~al.}(2015)Ihrig, Wieferink, Zhang, Ropo, Ren, Rinke,
  Scheffler, and Blum]{ihrig2015accurate}
Ihrig,~A.~C.; Wieferink,~J.; Zhang,~I.~Y.; Ropo,~M.; Ren,~X.; Rinke,~P.;
  Scheffler,~M.; Blum,~V. Accurate localized resolution of identity approach
  for linear-scaling hybrid density functionals and for many-body perturbation
  theory. \emph{New Journal of Physics} \textbf{2015}, \emph{17}, 093020\relax
\mciteBstWouldAddEndPuncttrue
\mciteSetBstMidEndSepPunct{\mcitedefaultmidpunct}
{\mcitedefaultendpunct}{\mcitedefaultseppunct}\relax
\EndOfBibitem
\bibitem[Yousaf and Peterson(2009)Yousaf, and Peterson]{yousaf2009optimized}
Yousaf,~K.~E.; Peterson,~K.~A. Optimized complementary auxiliary basis sets for
  explicitly correlated methods: aug-cc-pVnZ orbital basis sets. \emph{Chemical
  Physics Letters} \textbf{2009}, \emph{476}, 303--307\relax
\mciteBstWouldAddEndPuncttrue
\mciteSetBstMidEndSepPunct{\mcitedefaultmidpunct}
{\mcitedefaultendpunct}{\mcitedefaultseppunct}\relax
\EndOfBibitem
\bibitem[Jiang and Engel(2005)Jiang, and Engel]{Jiang/Engel:2005}
Jiang,~H.; Engel,~E. {Second-order Kohn-Sham perturbation theory: Correlation
  potential for atoms in a cavity}. \emph{The Journal of Chemical Physics}
  \textbf{2005}, \emph{123}, 224102\relax
\mciteBstWouldAddEndPuncttrue
\mciteSetBstMidEndSepPunct{\mcitedefaultmidpunct}
{\mcitedefaultendpunct}{\mcitedefaultseppunct}\relax
\EndOfBibitem
\bibitem[Jiang and Engel(2007)Jiang, and Engel]{jiang2007random}
Jiang,~H.; Engel,~E. Random-phase-approximation-based correlation energy
  functionals: Benchmark results for atoms. \emph{The Journal of chemical
  physics} \textbf{2007}, \emph{127}, 184108\relax
\mciteBstWouldAddEndPuncttrue
\mciteSetBstMidEndSepPunct{\mcitedefaultmidpunct}
{\mcitedefaultendpunct}{\mcitedefaultseppunct}\relax
\EndOfBibitem
\bibitem[Eshuis \latin{et~al.}(2010)Eshuis, Yarkony, and
  Furche]{eshuis2010fast}
Eshuis,~H.; Yarkony,~J.; Furche,~F. Fast computation of molecular random phase
  approximation correlation energies using resolution of the identity and
  imaginary frequency integration. \emph{The Journal of chemical physics}
  \textbf{2010}, \emph{132}, 234114\relax
\mciteBstWouldAddEndPuncttrue
\mciteSetBstMidEndSepPunct{\mcitedefaultmidpunct}
{\mcitedefaultendpunct}{\mcitedefaultseppunct}\relax
\EndOfBibitem
\bibitem[Booth(2006)]{booth2006power}
Booth,~T.~E. Power iteration method for the several largest eigenvalues and
  eigenfunctions. \emph{Nuclear science and engineering} \textbf{2006},
  \emph{154}, 48--62\relax
\mciteBstWouldAddEndPuncttrue
\mciteSetBstMidEndSepPunct{\mcitedefaultmidpunct}
{\mcitedefaultendpunct}{\mcitedefaultseppunct}\relax
\EndOfBibitem
\bibitem[Schwartz(1962)]{Schwartz:1962}
Schwartz,~C. Importance of Angular Correlations between Atomic Electrons.
  \emph{Phys. Rev.} \textbf{1962}, \emph{126}, 1015\relax
\mciteBstWouldAddEndPuncttrue
\mciteSetBstMidEndSepPunct{\mcitedefaultmidpunct}
{\mcitedefaultendpunct}{\mcitedefaultseppunct}\relax
\EndOfBibitem
\bibitem[Hill(1985)]{Hill:1985}
Hill,~R.~N. Rates of convergence and error estimation formulas for the
  Rayleigh-Ritz variational method. \emph{J. Chem. Phys.} \textbf{1985},
  \emph{83}, 1173\relax
\mciteBstWouldAddEndPuncttrue
\mciteSetBstMidEndSepPunct{\mcitedefaultmidpunct}
{\mcitedefaultendpunct}{\mcitedefaultseppunct}\relax
\EndOfBibitem
\end{mcitethebibliography}
